# Tailoring Reflectionless Complex Media for Non-Abelian Braiding of Acoustic Modes

Hongkuan Zhang[1], Guancong Ma[1†]

[1]Department of Physics, Hong Kong Baptist University, Kowloon Tong, Hong Kong, China

[†]Email: phgcma@hkbu.edu.hk

**Abstract**

Multiple scattering of sound and light can be tailored for diverse applications. Despite the great progress enabled by technologies such as time-reversal propagation and wavefront shaping, the full control of the transmission matrix remains a significant challenge. In this work, we propose a multi-scattering-based approach to design reflectionless complex media with an arbitrary unitary transmission matrix. As such, the perfect transmission of waves through such a medium performs a unitary operation. Based on this principle, we experimentally demonstrated braiding of multiple waveguide modes in an acoustic waveguide via multiple scattering and showed non-Abelian characteristics arising from the concatenation of distinct complex media. Furthermore, we show that the principle can be extended for realizing arbitrary unitary operations beyond braiding. Our scheme uses generalized Wigner-Smith operators to design the optimal acoustic complex media with near-arbitrary targeted functionalities. The scheme is generally applicable beyond acoustics, with broad implications to other wave types. Our results demonstrate unprecedented control over multiple-scattering waves and establish complex media as a viable route to modal control and operations on compact platforms, providing a new design paradigm for multimode, reconfigurable wave-based devices with potential applications in multiplexed communication, imaging, and the manipulation of quantum waves.

## 1. Introduction

Waves are multiply scattered in complex media. Although multiple scattering is generally expected to scramble the wavefront and destroy phase coherence [1], it can be harnessed for diverse functionalities that are inaccessible in "clean" wave systems [2–4]. For example, time-reversal refocusing of waves offers improved robustness and resolution in complex media [5–7]. Wavefront shaping can achieve focusing and imaging [8–10], deep targeted energy delivery [11,12], flexible communication control [13,14], and wave-based computation [15–17] through complex media. These phenomena hinge on the rich degrees of freedom of multiple scattering waves, which are embedded in a high-dimensional, richly structured scattering matrix (S-matrix). However, the generic S-matrix has limited open transmission channels [2,18,19], which means that, for general input wavefronts, the energy transmission coefficient is rather small, and a significant portion of the incident wave energy is reflected or absorbed. The less-than-perfect transmission is undesirable for wave control. A recent breakthrough reveals that by tailoring the coupling between two specimens of complementary complex media, all transmission channels can be fully opened, leading to perfect transmission of wave energy for arbitrary incident wavefronts [20,21]. However, although the outgoing wave carries all incident energy, the wavefronts are still scrambled, so the media appear translucent instead of transparent. In other words, the scheme still has almost no control over the waveform in each transmission channel. In this work, we develop and experimentally demonstrate a scheme for designing reflectionless complex media with an arbitrarily tailorable T-matrix. We show that such media can be used for realizing wave-based unitary operations. As a striking example, we have experimentally realized non-Abelian braiding of acoustic waveguide modes by multiple scattering—a functionality previously attainable only through delicate adiabatic evolutions in classical waves [22–25]. Our concept is based on the generalized Wigner-Smith operator (GWSO) [20,26–28], which guides a smart procedure for generating the desired configurations of acoustic complex media. By precisely controlling multiple waveguide modes, we implement the three generators of the braid group $B_4$ and directly observe the contrast between commutative and non-commutative braiding sequences. This braiding protocol fundamentally departs from prior realizations that rely on adiabatic evolutions in slowly varying structures [22,23], as it exploits multiple scattering to synthesize any unitary operation; for acoustics, this unlocks a robust and reconfigurable pathway to non-Abelian operations, where disorder becomes a resource rather than an obstacle. In addition, we demonstrate several distinct unitary operations, including two logical NOT gates and the fast Fourier transform, highlighting the generality and scalability of the method. Our work establishes complex media as a viable route to arbitrary modal control and operations on compact platforms, providing a new design paradigm for multi-mode, reconfigurable wave-based devices. Potential applications include multiplexed communication, imaging, and the manipulation of quantum waves.

## 2. Results

### 2.1. T-matrix for unitary operations

The complex medium consists of a collection of identical metallic cylinders placed in a two-dimensional (2D) rectangular multi-mode acoustic waveguide with air being the background (mass density $\rho_0 = 1.2\ \mathrm{kg/m^3}$, speed of sound $c_0 = 345\ \mathrm{m/s}$). The waveguide has a width of 20 cm. It supports four guiding modes at the operating frequency of 3.2 kHz. The cylinders have a radius of 2 cm (~1/5 the wavelength in air) and are considered sound-hard boundaries. Obviously, the scattering of an incident sound wave inside the waveguide is determined by the sizes, number, and positions of the cylinders. For simplicity, we establish a series of "scattering sections," each comprised of 80 cylinders. Each scattering section spans an area of $35 \times 20\ \mathrm{cm^2}$, and is designated to fulfill one function, as will be described later. The incident waves $|\psi_{\mathrm{in}}\rangle$ and outgoing waves $|\psi_{\mathrm{out}}\rangle$ of a scattering section connect as $|\psi_{\mathrm{out}}\rangle = \boldsymbol{S}|\psi_{\mathrm{in}}\rangle$, where $\boldsymbol{S}$ is the S-matrix. To establish the concept principle, we first neglect the intrinsic dissipation of the waveguide so the system respects time-reversal symmetry and obeys reciprocity under which $\boldsymbol{S}$ is unitary and symmetric, i.e., $\boldsymbol{S}^{\dagger}\boldsymbol{S} = \boldsymbol{I}$ and $\boldsymbol{S} = \boldsymbol{S}^{\mathrm{T}}$. The S-matrix can be expressed as

$$\boldsymbol{S} = \begin{pmatrix} \boldsymbol{R} & \boldsymbol{T}' \\ \boldsymbol{T} & \boldsymbol{R}' \end{pmatrix}, \tag{1}$$

where $\boldsymbol{R}$ and $\boldsymbol{R}'$ are the reflection matrices (R-matrices) for waves entering from the left and right sides of the medium, respectively, and $\boldsymbol{T}$ is the T-matrix. Note that $\boldsymbol{T}' = \boldsymbol{T}^{\mathrm{T}}$ due to reciprocity. In this paper, the S-matrix is expressed using the channel basis consisting of the four waveguide modes at 3.2 kHz, so $\boldsymbol{S}$ is $8 \times 8$ in dimensions.

Depending on the target functionality, the full medium may comprise $K$ serially concatenated scattering sections. The total S-matrix is given by $\boldsymbol{S}_{\mathrm{total}} = \boldsymbol{S}_1 \star \boldsymbol{S}_2 \star \cdots \boldsymbol{S}_{K-1} \star \boldsymbol{S}_K$, where $\star$ denotes the Redheffer star product, which is a complicated nonlinear operation[29]. However, when $\boldsymbol{R} = \boldsymbol{R}' = \boldsymbol{0}$, all wave energy is distributed in the T-matrix, such that $\boldsymbol{T}_n^{\dagger}\boldsymbol{T}_n = \boldsymbol{1}$, i.e., all T-matrices are unitary (see Appendix A for details). Importantly, the T-matrices now follow matrix multiplication, i.e.,

$$\boldsymbol{T}_{\mathrm{total}} = \boldsymbol{T}_K \cdots \boldsymbol{T}_2 \boldsymbol{T}_1. \tag{2}$$

In other words, the reflectionless propagation of sound through a series of scattering sections essentially performs linear unitary operations on the incident sound.

### 2.2. Braiding of waveguide modes

Braiding describes the ordered exchange of multiple neighboring "strands" [30]. It has profound applications in quantum physics owing to the relevance in quantum computation [31]. Recent studies have revealed that braiding can also be realized in classical wave systems [22,23,32–34], opening new routes for wave control. To show its relevance to our multiple scattering system, here we briefly introduce its mathematical construction. Consider the braiding of four strands, described by the Artin's braid group $B_4$

$$B_4 \coloneqq \{\sigma_{1,2,3} \mid \sigma_1\sigma_2\sigma_1 = \sigma_2\sigma_1\sigma_2, \sigma_3\sigma_2\sigma_3 = \sigma_2\sigma_3\sigma_2, \sigma_1\sigma_3 = \sigma_3\sigma_1\}, \tag{3}$$

where $\sigma_{1,2,3}$ are braid generators describing the exchange of the first and second, the second and third, and the third and fourth strands. Note that one needs to distinguish over-cross and under-cross of the strands, so one possible matrix representation is

$$\sigma_1(\eta) \to \begin{pmatrix} Y & 0 & 0 \\ 0 & 1 & 0 \\ 0 & 0 & 1 \end{pmatrix}, \sigma_2(\eta) \to \begin{pmatrix} 1 & 0 & 0 \\ 0 & Y & 0 \\ 0 & 0 & 1 \end{pmatrix}, \sigma_3(\eta) \to \begin{pmatrix} 1 & 0 & 0 \\ 0 & 1 & 0 \\ 0 & 0 & Y \end{pmatrix}, \tag{4}$$

where $Y(\eta) = \begin{pmatrix} 0 & e^{\mathrm{i}\eta} \\ -e^{-\mathrm{i}\eta} & 0 \end{pmatrix}$ with $\eta$ being a phase factor is an anti-symmetric, traceless $U(2)$ matrix. Apparently, $\sigma_{1,2,3}$ are unitary matrices. Also, it is straightforward to check that $B_4$ in a non-Abelian group with non-commutative elements, e.g., $\sigma_1\sigma_2 \neq \sigma_2\sigma_1$, $\sigma_3\sigma_2 \neq \sigma_2\sigma_3$.

The representation [Eqs. (4)] suggests that we can implement the braiding by engineering the proper T-matrix. In this context, the multiple scattering coverts the composition of acoustic waveguide modes at the input to a specific set of modes at the output. In other words, the desired braiding effect can be achieved by obtaining a medium with an S-matrix whose T-matrices are identical to those in Eqs. (4) and R-matrices vanish, i.e., $\boldsymbol{S}_{\mathrm{obj}} = \begin{pmatrix} \boldsymbol{0} & \boldsymbol{T}_{\mathrm{obj}}^{\mathrm{T}} \\ \boldsymbol{T}_{\mathrm{obj}} & \boldsymbol{0} \end{pmatrix}$ with $\boldsymbol{T}_{\mathrm{obj}} = \sigma_1$, $\sigma_2$ and $\sigma_3$. Taking $\boldsymbol{T}_{\mathrm{obj}} = \sigma_1$ as an example, in our acoustic setting, the desired medium should (1) be reflectionless; (2) fully convert the incidence in the first (second)-order guiding mode to the second (first)-order guiding mode at the output, while preserving the third and fourth-order modes; and (3) impart a phase delay (advance) of $\eta$ to the first (second)-order mode at the output, with the third and fourth-order modes are in phase. In previous works of wave-based braiding, $\eta$ was restricted to zero or $\pi$ because of the fundamental limitation in the system's underlying topology. Such a limitation does not exist in our scheme that is based on S-matrices, and it is possible to realize braiding with an arbitrary $\eta$. To unambiguously demonstrate this advantage and without the loss of generality, we choose $\eta = \pi/6$ in our experiment (This value is arbitrary and the method is not specific to any particular $\eta$). Furthermore, if the T-matrices of such two S-matrices correspond to the braiding operations $\sigma_i$ and $\sigma_j$, then the output waveguide mode is given by $\sigma_j \sigma_i |\psi_{\text{in}}\rangle$ (as shown in figure 1).

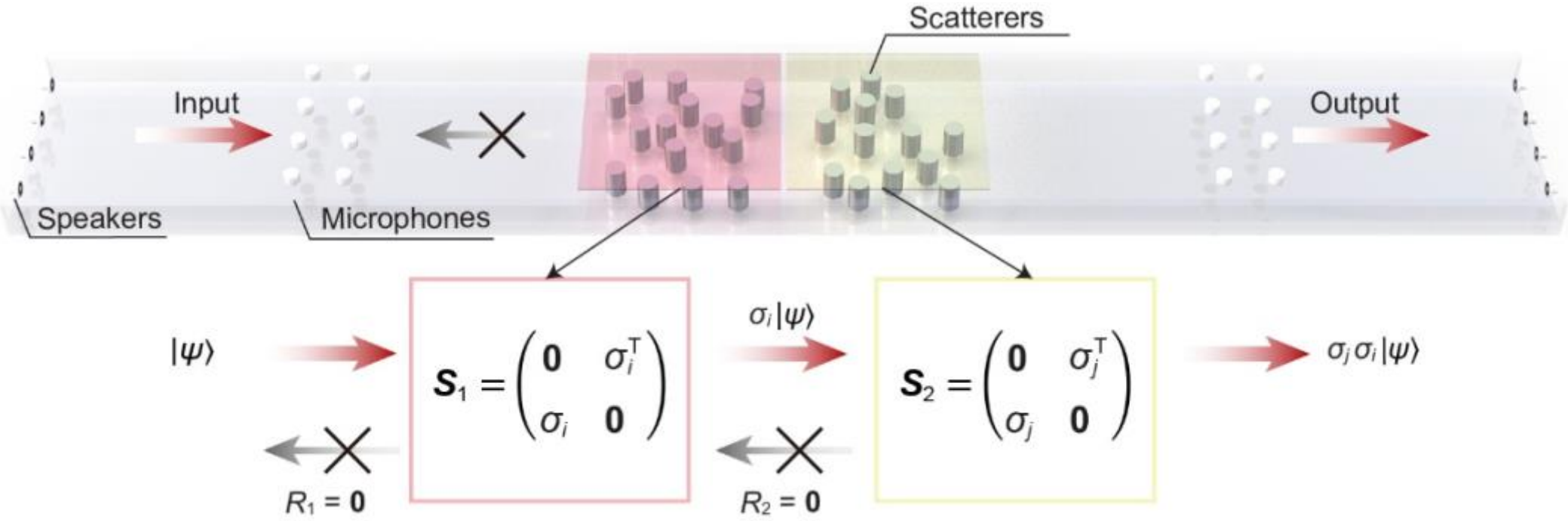

**Figure 1.** Schematic illustration of waveguide mode braiding enabled by multiple scattering media. The arrangement of cylindrical scatterers embedded in the waveguide act as the physical implementations of the operators $\sigma_i$ and $\sigma_j$. By arranging these scattering units in sequence, the system performs controlled mode conversion and wavefront modulation, enabling the braiding of guided modes.

### 2.3. Designing multiple scattering medium by GWSO

To achieve the said functionalities, we consider the S-matrix as a function of all positions of $M = 80$ cylindrical scatterers, i.e., $\boldsymbol{S}(\boldsymbol{r})$ with $\boldsymbol{r} = (x_1, x_2, \ldots, x_M, y_1, y_2, \ldots, y_M)$. These parameters form a smooth $2M$-dim manifold, on which a smooth evolution path connects the arbitrary $\boldsymbol{S}(\boldsymbol{r})$ to the desired $\boldsymbol{S}_{\text{obj}}$. Their proximity can be measured by using the squared Hilbert-Schmidt norm,

$$g(\boldsymbol{r}) = 1 - \frac{1}{4N^2} \left| \mathrm{Tr}\left( \boldsymbol{S}_{\text{obj}}^{\dagger} \boldsymbol{S}(\boldsymbol{r}) \right) \right|^2, \tag{5}$$

where Tr and † denote trace and conjugate transpose, respectively. Here, $N = 4$ denotes the number of channels. Basically, Eq. (5) maps the $2M$-dim parametric manifold $\boldsymbol{S}(\boldsymbol{r})$ to a single scalar $g(\boldsymbol{r}) \in [0, 1]$, which serves as a projection of $\boldsymbol{S}(\boldsymbol{r})$ onto $\boldsymbol{S}_{\text{obj}}$ and thus quantifies their similarity. The limit $g(\boldsymbol{r}) \to 0$ is our target, which indicates that the two matrices are identical up to a global phase factor $\phi$, i.e., $\boldsymbol{S}(\boldsymbol{r}) = e^{\mathrm{i}\phi}\boldsymbol{S}_{\mathrm{obj}}$. In practice, we minimize $g(\boldsymbol{r})$ using a gradient-descent optimization, where moving the scatterers amounts to applying a generalized "force" that drives $\boldsymbol{S}(\boldsymbol{r})$ along a smooth trajectory with a slope

$$\frac{\partial g(\boldsymbol{r})}{\partial r_m} = -\frac{1}{2N^2}\mathrm{Re}\left[\mathrm{Tr}\big(\boldsymbol{S}_{\mathrm{obj}}\boldsymbol{S}^{\dagger}(\boldsymbol{r})\big)\,\mathrm{Tr}\left(\boldsymbol{S}_{\mathrm{obj}}^{\dagger}\frac{\partial \boldsymbol{S}(\boldsymbol{r})}{\partial r_m}\right)\right]. \tag{6}$$

In Eq. (6), $\boldsymbol{S}(\boldsymbol{r})$ can be directly obtained with simulation or measurement so the only unknown term is the gradient of S-matrix $\frac{\partial \boldsymbol{S}(\boldsymbol{r})}{\partial r_m}$, where $r_m$ denotes the $m$-th component of the position vector $\boldsymbol{r}$ of the scatterers. We substitute this term by the GWSO $\boldsymbol{Q}_{r_m}$, which generates a change in $\boldsymbol{S}$ under a small variation of $r_m$, i.e., $\boldsymbol{S}(r_m + \Delta r_m) = \boldsymbol{S}(r_m)e^{\mathrm{i}\boldsymbol{Q}_{r_m}\Delta r_m}$, so

$$\frac{\partial \boldsymbol{S}(\boldsymbol{r})}{\partial r_m} = \mathrm{i}\boldsymbol{S}(\boldsymbol{r})\boldsymbol{Q}_{r_m}. \tag{7}$$

In our acoustic setting, $\boldsymbol{Q}_{r_m}$ is related to the total acoustic radiation force on the $m$-th scatterer, so it can be numerically evaluated [28,35] (see Appendix E for details). As such, we have all the components to build a gradient-descent optimization scheme to find the optimal position for each scatterer.

### 2.4. Experimental realization of non-Abelian braiding

We start with the 80 cylinders in a nearly random configuration. An example is shown in figure 2(A1). In this configuration, the corresponding T- and R-matrices are nearly random matrices, as shown in figure 2(A2). We further generate 100 independent random configurations, and their statistical analysis is shown in figure 2(A3). We note that the eigenvalue distribution of the energy transmission $\boldsymbol{T}^{\dagger}\boldsymbol{T}$ is close to a bimodal shape but the peak for unity transmission is missing. From these statistics, the scattering mean free path is estimated as $\ell_{\mathrm{sc}} = 2.8$ cm. Given that the average transmission rate across random configurations is about $\langle T\rangle = \langle \mathrm{Tr}\,\boldsymbol{T}^{\dagger}\boldsymbol{T}\rangle/N = 0.158$, the transport mean free path is estimated to be $\ell_{\mathrm{tr}} \simeq \langle T\rangle\ell = 5.5$ cm with $\ell = 35$ cm being the length of the scattering section. This places the system at $\ell \approx 6.4\ell_{\mathrm{tr}}$, beyond the nominal diffusive region $1 \lesssim \ell \lesssim N\ell_{\mathrm{tr}}$ (with number of channels $N = 4$ in this work) and is thus slightly more scattering than the ideally diffusive regime, which explains why the right peak of the bimodal distribution is not seen (see Appendix B for details). The results show that the scattering section is opaque. We then applied the optimization procedure described above with $\boldsymbol{S}_{\mathrm{obj}} = e^{\mathrm{i}\phi}\begin{pmatrix} 0 & \sigma_1^{\mathrm{T}}(\eta) \\ \sigma_1(\eta) & 0 \end{pmatrix}$ and $\eta = \pi/6$. The scatterer positions were iteratively adjusted until $g(\boldsymbol{r})$ dropped below $10^{-4}$. Figure 2(B1,

C1, D1) shows the optimized configurations of the selected scatterer arrangements in the experiment. The experimentally measured S-matrices are normalized by singular values to recover the lossless S-matrices. Loss manifests as an imaginary-frequency shift of the S-matrix, which rescales the singular values while leaving the phases unchanged (Detailed discussion on the matter of loss is presented in the Appendix G and SI section S5). For clarity, only normalized S-matrices are shown in the main text and unnormalized data are provided in SI section S5. In figure 2(B2), it is seen that the reflection is eliminated so R-matrix is near zero, and the T-matrix is exactly $\sigma_1(\pi/6)$ up to a global phase (here, $\phi_1 \cong -0.62\pi$). Similarly, $\sigma_2(\pi/6)$ and $\sigma_3(\pi/6)$ are also achieved, as shown in figure 2(C2, D2), respectively. The corresponding global phases are $\phi_2 \cong 0.75\pi$ and $\phi_3 \cong 0.54\pi$. In addition, we further performed 40 independent optimizations targeting $\sigma_1$, $\sigma_2$ and $\sigma_3$, and plotted the energy transmission and reflection spectra in figure 2(B3, C3, D3). In the pre-optimized configuration, the average transmission and reflection were both close to 0.5. The results demonstrate that, near the target frequency of 3.2 kHz, the optimized medium is completely transparent to wave energy.

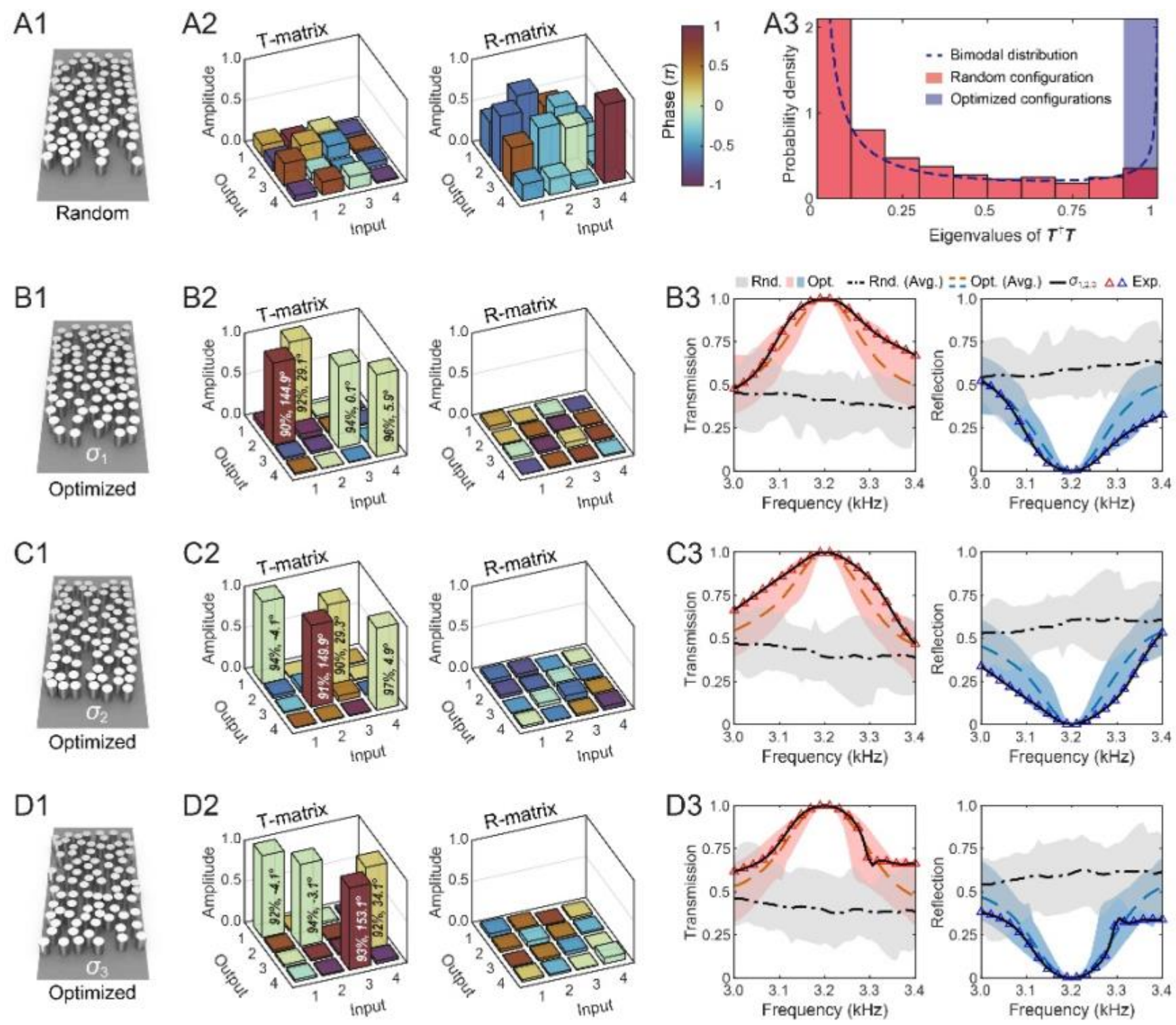

**Figure 2.** The optimized complex media realizing the three braid generators in their T-matrices. (A) A pre-optimized, random configuration: (A1) scatterer layout, (A2) corresponding T- and R-matrices, and (A3) transmission eigenvalue distribution over 40 random realizations, showing a bimodal distribution. (B-D)

Optimized configurations for $\sigma_1$, $\sigma_2$, and $\sigma_3$, with layouts shown in (B1, C1, D1). (B2-D2) Experimentally measured T- and R-matrices. (B3-D3) Transmission and reflection spectra: shaded bands denote the range over 40 independent realizations (gray, random; red/blue, optimized). Black dash-dot (random) and dashed (optimized) curves indicate ensemble averages. Solid black curves correspond to the specific layouts in (B1-D1), and triangles mark experimental data. All experimental spectra are normalized to recover the lossless responses.

To further demonstrate the effect of mode braiding, we raster-scanned the output fields in the waveguide using a microphone mounted on a 2D translational stage. The effect of $\sigma_1$ on the waveguide modes is shown in figure 3(A1-A2). It is seen that the optimized medium totally converts the first (second)-order waveguide mode to the second (first)-order mode, whereas it is totally transparent to the third and fourth-order modes. Specifically, incident waves in channels $|1\rangle$, $|2\rangle$, $|3\rangle$, and $|4\rangle$ are scattered into $e^{\mathrm{i}\pi/6}|2\rangle$, $-e^{-\mathrm{i}\pi/6}|1\rangle$, $|3\rangle$ and $|4\rangle$, respectively. The wavefield measurements further confirm the results [right-hand insets in figure 3(A2)]. The phases of the output modes are shown in figure 3(A3), wherein $\eta = \pi/6$ is clearly seen. The other two generators $\sigma_2$ and $\sigma_3$ exhibit the intended functionalities, as shown in figure 3(B, C).

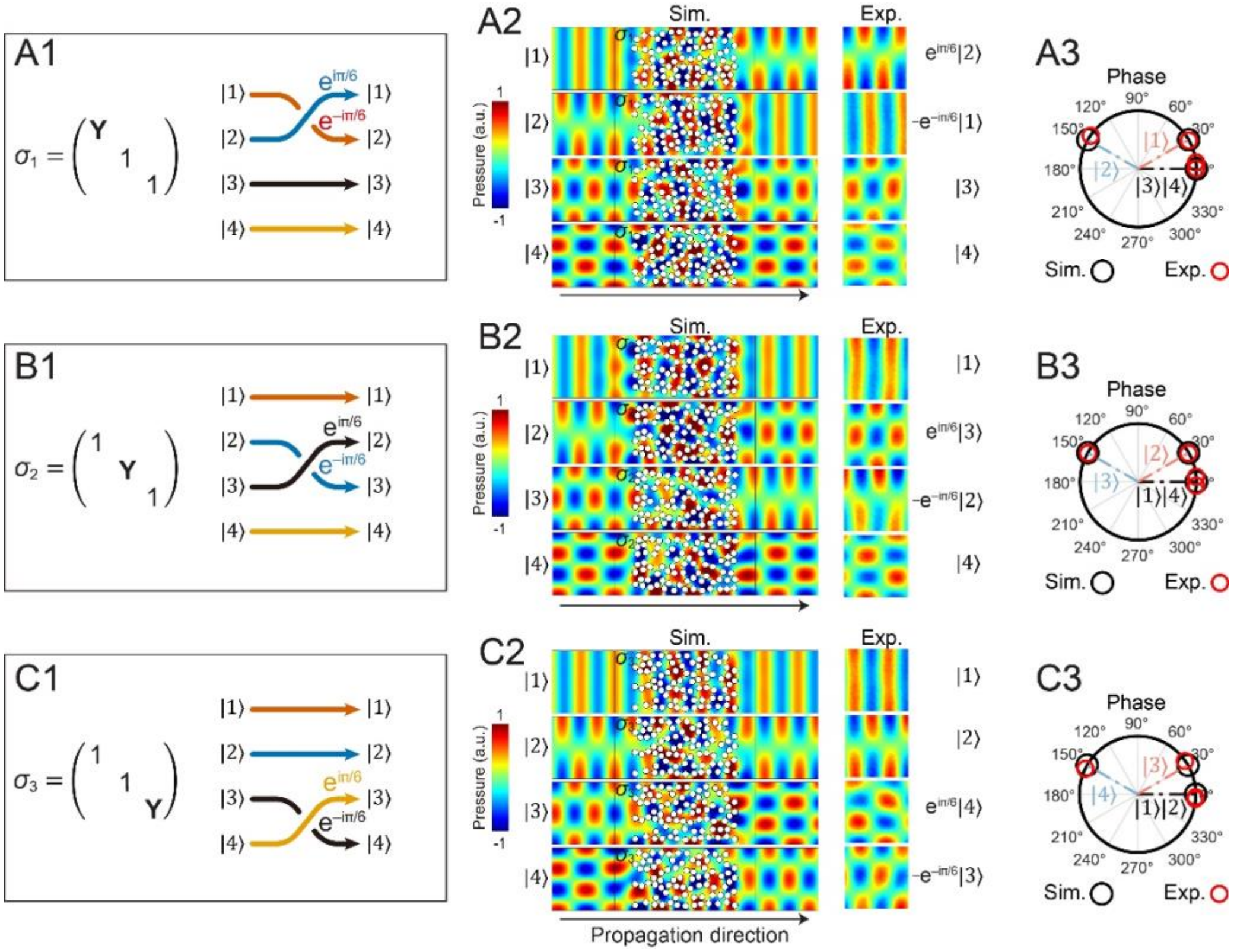

**Figure 3.** Braiding effects on waveguide modes induced by complex media corresponding to the three generators. (A1-C1) Braiding diagram of $\sigma_1$, $\sigma_2$ and $\sigma_3$. (A2-C2) Simulated (left) and experimental (right) field distributions for the three generators. Pure input modes are launched from the left; after propagation through the complex medium, the expected braided output modes are obtained. (A3-C3) Output-mode phase distribution after removing the global phase: $\sigma_1$: modes $|3\rangle$ and $|4\rangle$ remain at 0, while modes $|1\rangle$ and $|2\rangle$ acquire additional phases of 30° and 150°, respectively. $\sigma_2$: modes $|1\rangle$ and $|4\rangle$ remain at 0, while modes $|2\rangle$ and $|3\rangle$ acquire phases of 30° and 150°, respectively. $\sigma_3$: modes $|1\rangle$ and $|2\rangle$ remain at 0, while modes $|3\rangle$ and $|4\rangle$ acquire phases of 30° and 150°, respectively.

### 2.3. Demonstration of non-Abelian characteristics

The braiding group $B_4$ is a non-Abelian group characterized by the relation $[\sigma_i, \sigma_j] \neq 0$ for $|i-j| = 1$, i.e., sequential braid operations involving a shared mode do not commute. To verify that such a relation can be realized in our system, we stack two different scattering sections that respectively performs $\sigma_1$ and $\sigma_2$. Since the two operations do not commute, i.e., $[\sigma_1, \sigma_2] \neq 0$, their order affects the mode of the output. As shown in figure 4(A1, B1), when the scattering sections are placed in the order $\sigma_1 \rightarrow \sigma_2$, incident waves in channels $|1\rangle$, $|2\rangle$, $|3\rangle$, and $|4\rangle$ are converted to $e^{-\mathrm{i}\pi/3}|3\rangle$, $e^{\mathrm{i}\pi/6}|1\rangle$, $e^{\mathrm{i}\pi/6}|2\rangle$, and $|4\rangle$, respectively. Conversely, when the scattering sections are

arranged in the order $\sigma_2 \rightarrow \sigma_1$, the incident waves in channels $|1\rangle$, $|2\rangle$, $|3\rangle$, and $|4\rangle$ are scattered into $e^{\mathrm{i}5\pi/6}|2\rangle$, $e^{\mathrm{i}5\pi/6}|3\rangle$, $e^{\mathrm{i}\pi/3}|1\rangle$, and $|4\rangle$. These effects are observed in numerical simulations and are validated in experimental measurements, as shown in figure 4(A2, B2). The T-matrices further confirm the non-Abelian characteristics [figure 4(A3, B3)]. The results clearly demonstrate the impact of operation order on the waveguide modes and phases.

Next, we consider the cascade of scattering sections for $\sigma_1$ and $\sigma_3$. These two operators do commute, so the order of operations does not affect the braiding of waveguide modes. As shown in figure 4(C1-C3, D1-D3), both $\sigma_1\sigma_3$ and $\sigma_3\sigma_1$ convert incident waves in channels $|1\rangle$, $|2\rangle$, $|3\rangle$, and $|4\rangle$ to $e^{\mathrm{i}5\pi/6}|2\rangle$, $e^{\mathrm{i}\pi/6}|1\rangle$, $e^{\mathrm{i}5\pi/6}|4\rangle$, and $e^{\mathrm{i}\pi/6}|3\rangle$. Owing to the limited space in our lab, we numerically verified the commutative relations of three sequential braid generators, i.e., the Yang-Baxter relation, in SI section S2.

### 2.4. Realizations of other unitary operations

Our method can generate unitary operations beyond those from a braid group. To show this, we present two representative examples: the bitwise NOT (BNOT) and the controlled NOT (CNOT). We first encode the waveguide modes in binary form, namely $|1\rangle \mapsto |00\rangle$, $|2\rangle \mapsto |01\rangle$, $|3\rangle \mapsto |10\rangle$, $|4\rangle \mapsto |11\rangle$. The matrix representations of the two operations are

$$\mathrm{BNOT} \rightarrow \begin{pmatrix} 0 & 0 & 0 & 1 \\ 0 & 0 & 1 & 0 \\ 0 & 1 & 0 & 0 \\ 1 & 0 & 0 & 0 \end{pmatrix}, \mathrm{CNOT} \rightarrow \begin{pmatrix} 1 & 0 & 0 & 0 \\ 0 & 1 & 0 & 0 \\ 0 & 0 & 0 & 1 \\ 0 & 0 & 1 & 0 \end{pmatrix}. \tag{8}$$

Using the same optimization procedure, we obtained complex media that implement these two operators. Experimental measurements for BNOT are shown in figure 5(A): the transmission block clearly reveals the structure of the target matrix, while the reflection block is suppressed. The mode-conversion is observed in simulation and experiment, as shown in figure 5(B1, B2). Similar results for CNOT are shown in figure 5(C, D1-D2).

In principle, our approach can be used to realize arbitrary unitary matrices. As a further example, we numerically demonstrate using complex media for the discrete Fourier transform (DFT) of a temporal signal. By designing the T-matrix to match the DFT matrix, we can convert an input vector into its frequency-domain representation purely via multiple scattering. This shows that complex media can perform fundamental signal-processing tasks directly at the physical layer. A detailed description of the design and numerical results is given in SI Section S3.

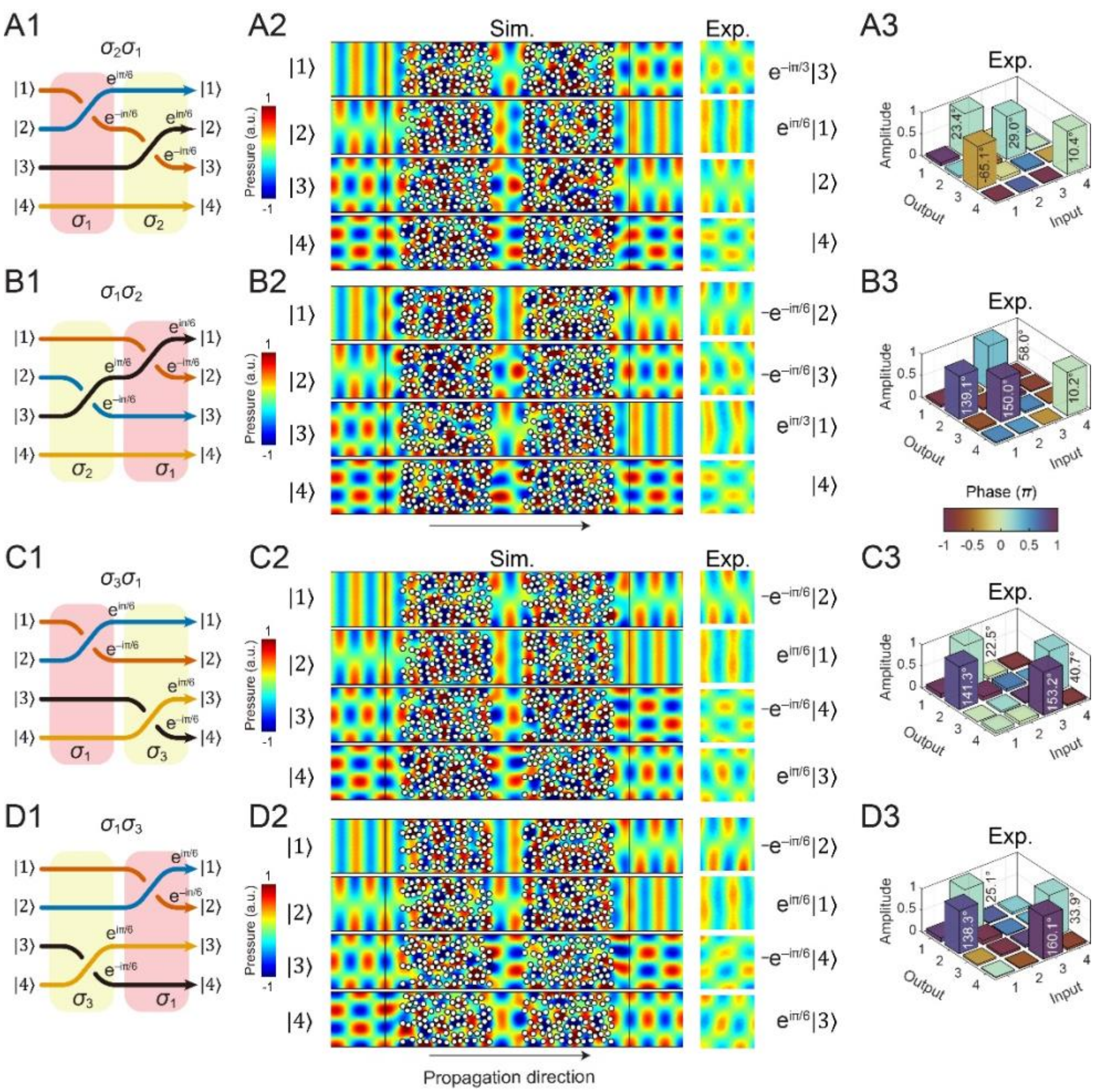

**Figure 4.** Concatenation of braid operations demonstrating non-Abelian characteristics. (A1, B1) Braiding effects produced by the sequences $\sigma_2\sigma_1$, and $\sigma_1\sigma_2$ (A2, B2). Numerical simulations and experimental results (right panels) show that $\sigma_1$ and $\sigma_2$ are non-commutative: the output waveguide modes depend on the ordering of the two operators. (A3, B3) Experimentally measured T-matrices of the combined complex media. (C1, D1) Braiding effects produced by the sequences $\sigma_3\sigma_1$ and $\sigma_1\sigma_3$. (C2, D2) Numerical simulations and experimental results (right panel) demonstrate that $\sigma_1$ and $\sigma_3$ commute: the output waveguide modes are independent of operator ordering. (C3, D3) Experimentally measured T-matrices of the corresponding combined complex media.

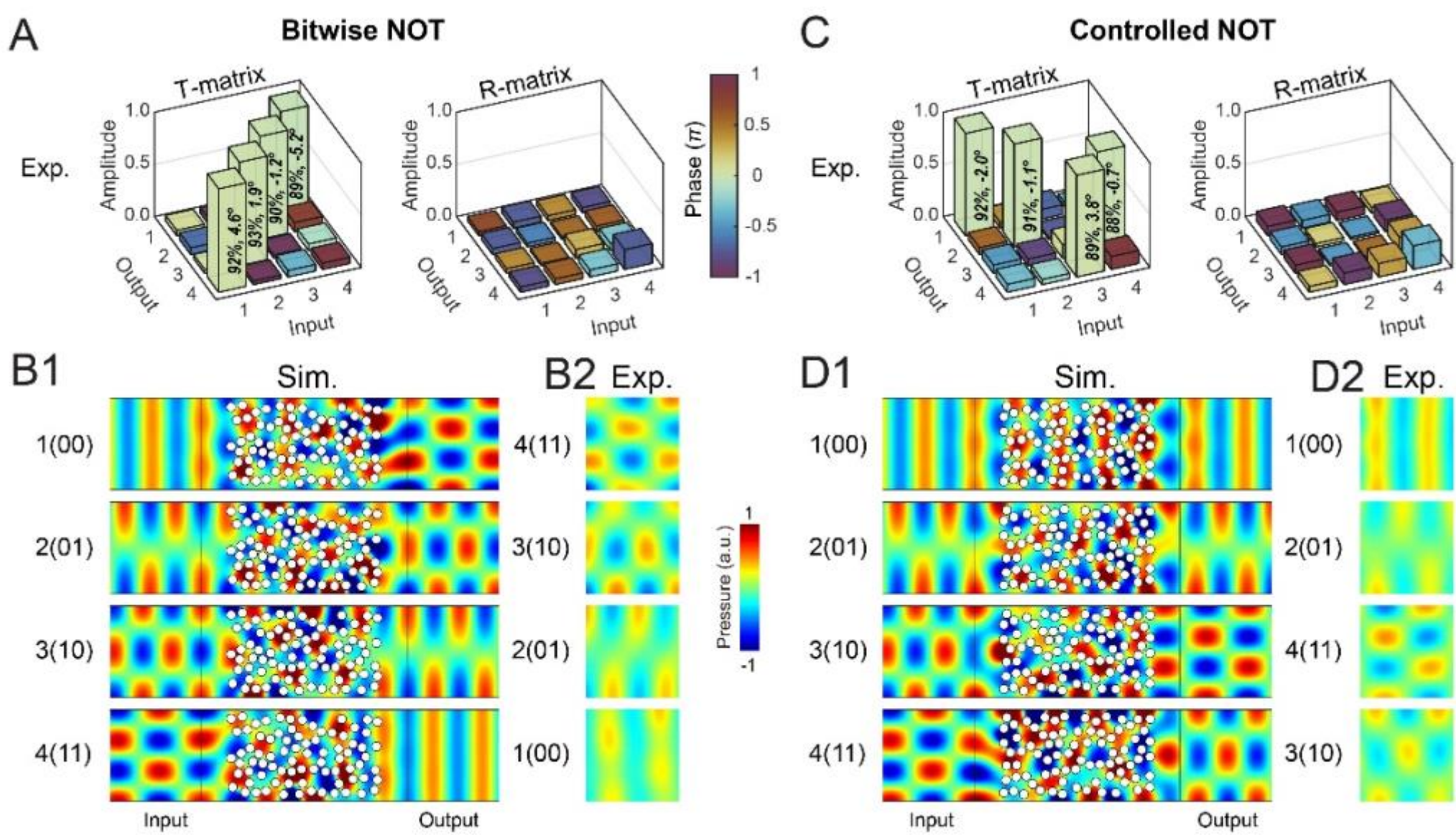

**Figure 5.** Implementation of the BNOT and CNOT operations. (A) Experimentally measured T- and R-matrices of the optimized medium realizing the BNOT gate. (B1-B2) Mode-conversion patterns for BNOT obtained from numerical simulations. (C) Experimentally measured T- and R-matrices of the optimized medium implementing the CNOT gate. (D1-D2) Mode-conversion patterns for CNOT from numerical simulations.

## 3. Discussion and conclusion

### 3.1. Discussion

Several limitations should be noted. First, the complex media were designed for a single frequency by the GWSO optimization. However, our investigation shows that the fidelity decays only quadratically with frequency detuning, so the media retain good functionality over a useful bandwidth. In our case, the fidelity is better than 0.9 over 160 Hz at the operation frequency of 3.2 kHz. Second, the acoustic loss enters through a purely imaginary frequency in the S-matrix. It only rescales the singular values linearly but does not introduce a global phase shift. As such, its impact on the functionalities can be easily compensated for by channel-wise singular value normalization. Third, the functionalities have reasonably good tolerances against errors in the scatterers' positions also lead to a quadratic fidelity decay. Our estimation shows that a fidelity well above 0.95 can be achieved via placement by hand with simple visual guidance. Detailed analyses on the bandwidth, loss, and position error are shown in Appendices G and H and SI section S5-S7.

Previous schemes of realizing braiding in classical wave [22,23,32–34] rely on adiabatic evolutions, which mandate the waves to smoothly evolve into the desired distributions. This approach requires very long structures (typically over 30 wavelengths for one braid operation) and delicate geometric details in the waveguides. In comparison, our scheme here realizes braiding by designing disordered multiple-scattering media whose T-matrices directly serve as the braid generators. This removes the need for adiabaticity, resulting in much thinner structures—the thickness of one scattering section is 35 cm or roughly 3.2 wavelengths. Our approach is also clearly distinct from non-adiabatic time-varying methods [36]. We also highlight that the scheme here can realize braiding with an arbitrary phase factor $\eta$, which is far more general compared to previous methods that can only achieve $\eta = 0$ or $\pi$ [22,23,34]. This opens the possibility of emulating anyonic exchange behavior in waves.

### 3.2. Conclusion and outlook

Our work demonstrates that the S-matrix of a complex medium can be directly and precisely tailored for unitary mode transformations by encoding the operation in the T-matrices. Such unitary transformations encompass diverse wave-based functionalities, including wave multiplexing [16,37], imaging [3,9], wavefield shaping [8,38], and quantum simulation [39]. Although the same can in principle also be realized by sequences of non-reflecting phase masks [40], our scheme that relies on complex media does not require precise multi-stage alignment and has much shorter thickness. Compared to metamaterials or metasurfaces, our scheme does not involve resonances in subwavelength scales [41,42] and, therefore, is less lossy and has a wider bandwidth. In addition, our method requires neither nonlinear effects nor active gain nor intricate microstructural designs. We also highlight that the GWSOs that are central to our work are a scalable and universal theoretical framework that is applicable to other systems such as microwaves, elastic waves, and optics.

Future investigations may focus on controlling more complex wavefields in open environments or including active programmability. Since non-unitary S- and T-matrices also commonly appear in multiple scattering media, it may be interesting to investigate them for wave-based non-unitary operations. Adaptation of more advanced computational methods, such as machine learning or artificial intelligence, may also be a fruitful direction.

### Data availability statement

The data that generate the results of this study are available from the corresponding authors upon request.

## Funding

This work was supported by the National Natural Science Foundation of China (NSFC, T2525002), the National Key R&D Program (2022YFA1404400), the Hong Kong Research Grants Council (RFS2223-2S01, 12301822, 12300925), and the Hong Kong Baptist University (RC-RSRG/23-24/SCI/01, RC-SFCRG/23-24/R2/SCI/12).

## Author contributions

G. M. initiated and supervised the project. H. Z. performed theoretical derivations, numerical simulations, and experimental measurements. Both authors analyzed the data and wrote the paper.

**Competing interests.** The authors declare no competing interests.

## Appendix

### A. S-matrix of cascaded scattering sections

For two cascaded scattering sections with S-matrices $\boldsymbol{S}_1$ and $\boldsymbol{S}_2$, the total S-matrix is given by the Redheffer star product[29], denoted as $\boldsymbol{S}_{\text{total}} = \boldsymbol{S}_1 \star \boldsymbol{S}_2$, whose blocks components are

$$\begin{aligned} \boldsymbol{R}_{\text{total}} &= \boldsymbol{R}_1 + \boldsymbol{T}_1'(\boldsymbol{I} - \boldsymbol{R}_2\boldsymbol{R}_1')^{-1}\boldsymbol{R}_2\boldsymbol{T}_1 \\ \boldsymbol{R}_{\text{total}}' &= \boldsymbol{R}_2' + \boldsymbol{T}_2(\boldsymbol{I} - \boldsymbol{R}_1'\boldsymbol{R}_2)^{-1}\boldsymbol{R}_1'\boldsymbol{T}_2' \\ \boldsymbol{T}_{\text{total}}' &= \boldsymbol{T}_1'(\boldsymbol{I} - \boldsymbol{R}_2\boldsymbol{R}_1')^{-1}\boldsymbol{T}_2' \\ \boldsymbol{T}_{\text{total}} &= \boldsymbol{T}_2(\boldsymbol{I} - \boldsymbol{R}_1'\boldsymbol{R}_2)^{-1}\boldsymbol{T}_1 \end{aligned}. \tag{9}$$

A notable simplification occurs when all reflection blocks are zero, i.e., $\boldsymbol{R}_1 = \boldsymbol{R}_1' = \boldsymbol{R}_2 = \boldsymbol{R}_2' = \boldsymbol{0}$, which gives

$$\boldsymbol{R}_{\text{total}} = \boldsymbol{R}_{\text{total}}' = \boldsymbol{0}, \qquad \boldsymbol{T}_{\text{total}}' = \boldsymbol{T}_1'\boldsymbol{T}_2', \qquad \boldsymbol{T}_{\text{total}} = \boldsymbol{T}_2\boldsymbol{T}_1. \tag{10}$$

Equations (10) indicate perfect transmission between the two sections. Importantly, the total T-matrix is obtained from the individual T-matrices through matrix multiplication. This result leads to Eq. (2), which is foundational to our work.

### B. Scattering and transport mean free path

**Scattering mean free path.** In the 2D case, the scattering mean free path is estimated as $\ell_{\mathrm{sc}} = \pi S/L$ [43,44], where $S$ is the area available for sound wave propagation (the area of the rectangular region minus the total area of all the circular scatterers, $\sim 0.04487\mathrm{m}^2$) and $L$ is the total length of all scattering boundaries (the total perimeter of the scatterers, $\sim 5.027\mathrm{m}$). This gave an approximate mean free path of $2.8\mathrm{cm}$.

**The transport mean free path.** In the waveguide, the transport mean free path can be estimated from random-matrix theory via $\langle T \rangle \simeq \ell_{\mathrm{tr}}/\ell$ [2,45], where $\ell = 35$ cm is the length of the scattering section and $\langle T \rangle$ is the average energy transmission. We compute energy transmission, $\frac{1}{N}\mathrm{Tr}\,\boldsymbol{T}^{\dagger}\boldsymbol{T}$, for all 100 initial random configurations and take the mean. With $\langle T \rangle = \langle \mathrm{Tr}\,\boldsymbol{T}^{\dagger}\boldsymbol{T} \rangle / N = 0.158$, the transport mean free path is estimated as $\ell_{\mathrm{tr}} \simeq \langle T \rangle \ell = 5.5$ cm, where $N$ is the number of channels. Note that the complex medium region spans approximately $6.4\ell_{\mathrm{tr}}$, which exceeds $N = 4$. The system is, therefore, more scattering than the diffusive regime. Consequently, the right peak of the bimodal distribution in figure 2A3 is not prominent [2].

### C. Experimental methods

We performed experiments in a 2D multimode acoustic waveguide and implemented a measurement protocol to reconstruct the channel-based S-matrix. We use 5 mm-thick aluminum alloy plates to build an air-filled waveguide. The length of the waveguide is $D = 2$ m, and the cross-section has a width of $W = 0.2$ m, and a height of $H = 0.03$ m. A complex medium comprised of aluminum alloy cylinders with a radius of 2 cm is placed at the optimized position in the middle of the waveguide. The scatterers are dispersed to a $35 \times 20\ \mathrm{cm}^2$ region inside the waveguide. When defining the S-matrix of a complex medium, an additional 5-cm air section is included on each side of the complex medium to mitigate the influence of near-field evanescent waves on the S-matrix evaluation. The operating frequency is set to 3.2 kHz, at which the waveguide supports $N = 4$ propagating modes that serve as the channels. At each end of the waveguide, 4 loudspeakers are installed. Between the scattering region and the two ends, a total of $4N = 16$ microphones are deployed for sound measurements. On both left and right sides, two parallel measurement lines are positioned with a spacing of 2.5 cm (which can avoid resonances within the frequency band of interest); each line contains 4 microphones spaced 5 cm apart, enabling detailed sampling of the internal sound field distribution. Signal generation and acquisition are implemented by a MOTU 16A audio interface with 16 inputs, 16 outputs, synchronized, which are programmable via MATLAB. The measured pressure fields are decomposed into waveguide modes to construct the S-matrix. In a homogeneous background (density $\rho \equiv \rho_0$ and sound speed $c \equiv c_0$), the acoustic pressure $P(x, y)$ satisfies the Helmholtz equation $\boldsymbol{\nabla}^2 P(x, y) + \frac{\omega^2}{c_0^2} P(x, y) = 0$ , whose eigenfunctions form a complete orthogonal basis of waveguide modes

$$
\begin{aligned}
P_n(x, y) &= \frac{\sqrt{2 - \delta_{1n}}}{\sqrt{W}} \frac{\cos k_n^y y}{\sqrt{k_n^x}} \left( a_n e^{-\mathrm{i} k_n^x x} + b_n e^{\mathrm{i} k_n^x x} \right) \qquad \text{for } x \leq 0 \\
P_n(x, y) &= \frac{\sqrt{2 - \delta_{1n}}}{\sqrt{W}} \frac{\cos k_n^y y}{\sqrt{k_n^x}} \left( c_n e^{-\mathrm{i} k_n^x (x-L)} + d_n e^{\mathrm{i} k_n^x (x-L)} \right) \quad \text{for } x \geq L
\end{aligned}
\quad , \tag{11}
$$

where transverse and propagative wavenumbers are $k_n^y = (n-1)\pi/W$ and $k_n^x = \sqrt{\omega^2/c_0^2 - \left(k_n^y\right)^2}$, respectively, with $n$ index the modes. All propagating modes are normalized by $1/\sqrt{k_n^x}$ such that they carry the same energy flux and thus serve as independent transport channels. The number of propagating modes is $N = 1 + \lfloor \omega W/(c_0 \pi) \rfloor$, where $\lfloor \cdot \rfloor$ denotes the floor function. The indices of the incident and outgoing channels are summarized into two vectors, $|\psi_{\mathrm{in}}\rangle = \left(a_1, \ldots, a_{N_{\mathrm{ch}}}, d_1, \ldots, d_{N_{\mathrm{ch}}}\right)^{\mathrm{T}}$, and $|\psi_{\mathrm{out}}\rangle = \left(b_1, \ldots, b_{N_{\mathrm{ch}}}, c_1, \ldots, c_{N_{\mathrm{ch}}}\right)^{\mathrm{T}}$, their relation is given by $|\psi_{\mathrm{out}}\rangle = \boldsymbol{S} |\psi_{\mathrm{in}}\rangle$, where $\boldsymbol{S}$ is the S-matrix. Each element $S_{mn}$ denotes the scattering coefficient from the $n$-th channel in the incidence to the $m$-th channel in the outgoing. The S-matrix can then be obtained from the measured data using a procedure detailed in SI section S1.

The sound field maps (e.g., results in Figures 3, 4) were obtained by a microphone mounted on a dual-axis translational stage, which raster-scanned a 20 cm × 20 cm region with a step size of 5 mm. To improve the signal-to-noise ratio, time-domain windowing and frequency-domain filtering are applied. This is followed by phase unwrapping and background subtraction to ensure robust reconstruction stability.

**D. Construction of the S-matrix**

The scattering fields are numerically simulated using COMSOL Multiphysics. The acoustic waveguide and the scatterers are modeled in 2D using the acoustic module, with the height dimension neglected. Perfectly matched layers are used at both ends of the $x$-axis to absorb outgoing waves such that the S-matrix can be correctly obtained. The acoustic pressure field $P(x,y)$ satisfies the generalized Helmholtz equation

$$\mathcal{L}P(x,y) = \rho(x,y)\nabla\cdot(\rho^{-1}(x,y)\nabla P(x,y)) + \frac{\omega^2}{c^2(x,y)}P(x,y) = 0, \tag{12}$$

where $\rho$ and $c$ are the density and sound speed, respectively, and $\omega$ is the angular frequency. The operator $\mathcal{L}$ explicitly incorporates the spatial distribution of inertia. The field data are exported to MATLAB via the COMSOL LiveLink, where the S-matrix and the GWSO are subsequently computed. Numerically, $\boldsymbol{S}$ is constructed by exciting each input channel individually and recording the resulting outgoing amplitudes [44].

### E. The generalized Wigner-Smith operator

The GWSO associated with variations in scatterer positions quantifies the resulting change in momentum, which corresponds directly to the acoustic radiation force[27,28,35]. Since the scattering originates from spatial inhomogeneities in material parameters, the evaluation of this force can ultimately be expressed as a surface integral of the radiation stress over the inclusion boundary. In acoustics, both density and compressibility contribute to the force, corresponding to momentum-flux and pressure-gradient terms, respectively[46]. Our derivations and computations are carried out under rigid-boundary conditions.

In this work, we formulate the GWSO within the framework of the Gateaux derivative[47–49]. which differs from prior variational and radiation-force approaches [28,35] and thus provides a complementary operator-level contribution. Consider an infinitesimal displacement of the $m$-th scatterer's boundary $\Gamma_m$, i.e., $\delta\boldsymbol{r}_m = \alpha\boldsymbol{e}_m$, where $\boldsymbol{e}_m$ denotes the unit vector in the direction of translation and $\alpha$ is an infinitesimal scalar. The corresponding change in the wave operator $\mathcal{L}$ due to this boundary shift is captured by its shape derivative $\frac{\partial\mathcal{L}}{\partial\alpha}$. The GWSO $\boldsymbol{Q}_{r_m}$ associated with displacement relates directly to the acoustic radiation force. Its matrix elements are given by[27]

$$\langle\psi_\mu|\boldsymbol{Q}_{r_m}|\psi_\nu\rangle = -\frac{1}{2}\left\langle P_{\psi_\mu}\middle|\frac{\partial\mathcal{L}}{\partial\alpha}\middle|P_{\psi_\nu}\right\rangle, \tag{13}$$

where, $|\psi_\mu\rangle$ is the $\mu$-th incoming waveguide mode, whereas $|P_{\psi_\mu}\rangle$ is the corresponding acoustic field inside the scattering region. From Eq. (13), the components of the GWSO for $m$-th scatterer reduce to integrals taken along its boundary $\Gamma_m$ (rigid boundary), as follows

$$\begin{pmatrix} Q_x^{\mu\nu} \\ Q_y^{\mu\nu} \end{pmatrix} = \frac{1}{2}\int_{\Gamma_m} \left(k_0^2 P_{\psi_\mu}^* P_{\psi_\nu} - \partial_\parallel P_{\psi_\mu}^* \partial_\parallel P_{\psi_\nu}\right) \begin{pmatrix} n_x \\ n_y \end{pmatrix} \mathrm{d}\Gamma_m \,, \tag{14}$$

where $\partial_\parallel = \boldsymbol{\nabla} - \boldsymbol{n} \cdot \boldsymbol{\nabla}$ is the tangential gradient, and $\boldsymbol{n} = (n_x, n_y)$ is the outward unit normal. This expression differs from its microwave counterpart, where radiation pressure arises solely from contrasts in the dielectric constant[20,27] and therefore contains no term involving the tangential gradient. A complete derivation is provided in SI section S4.

### F. Optimization algorithm

To realize the target matrix $\boldsymbol{S}_{\mathrm{obj}}$, we drive $\boldsymbol{S}(\boldsymbol{r})$ toward the desired response using gradient-based optimization. At each iteration, the positions of the cylinders are updated so that the resulting S-matrix progressively approaches the target $\boldsymbol{S}_{\mathrm{obj}}$. The update follows the gradient-descent rule, in which the cylinders move in the direction opposite to the gradient of the objective function. Let the scatterers' positions be collected in $\boldsymbol{r} = (x_1, x_2, \cdots, y_1, y_2, \cdots)^{\mathrm{T}}$, with the current iterate $\boldsymbol{r}_{(k)}$. The gradient of the objective $g(\boldsymbol{r})$ is evaluated using Eq. (6) as $\boldsymbol{\nabla} g = \left(\frac{\partial g}{\partial x_1}, \frac{\partial g}{\partial x_2}, \cdots, \frac{\partial g}{\partial y_1}, \frac{\partial g}{\partial y_2}, \cdots\right)^{\mathrm{T}}$, and the parameters are updated according to

$$\boldsymbol{r}_{(k+1)} = \boldsymbol{r}_{(k)} + \Delta\boldsymbol{r}_{(k)}, \text{with } \Delta\boldsymbol{r}_{(k)} = -\eta \frac{\boldsymbol{\nabla} g(\boldsymbol{r}_{(k)})}{\|\boldsymbol{\nabla} g(\boldsymbol{r}_{(k)})\|}. \tag{15}$$

The objective decreases approximately as

$$g(\boldsymbol{r}_{(k+1)}) \approx g(\boldsymbol{r}_{(k)}) - \eta \|\boldsymbol{\nabla} g(\boldsymbol{r}_{(k)})\|. \tag{16}$$

The initial learning rate $\eta$ is chosen such that the maximum displacement of each cylinder does not exceed 20% of its diameter. To stabilize convergence, we employ the YOGI optimizer[50], which is an ADAM-type method that accumulates locally weighted gradients. This smoothing improves the update direction and accelerates the convergence of $\boldsymbol{S}(\boldsymbol{r})$ toward the desired unitary transformation.

To prevent collisions between the cylinders, geometric constraints are enforced throughout the optimization. The initial configuration is generated by partitioning the scattering region into equally sized cells. Scatterers are placed randomly inside these cells, subject to small random perturbations, while ensuring no overlap occurs either between scatterers or with the waveguide boundaries. Since the number of cells exceeds the number of scatterers, sufficient randomness in the initial arrangement is guaranteed. During the optimization, if a displacement would cause a scatterer to intersect the boundary, only its tangential component is retained. If a displacement would result in an overlap between two scatterers, only the component perpendicular to the line connecting their centers is kept. Should overlap persist after this adjustment, the step size is halved iteratively until the constraint is satisfied and convergence is achieved.

Our simulations were performed on an Intel Xeon W3-2435 CPU (3.1 GHz, 8 cores, 128 GB RAM). A single COMSOL finite-element solve (~50,000 degrees of freedom) takes about 3~5 s; and extracting the S-matrix and Q-matrices via MATLAB takes ~1 s. A typical optimization converges within 200 generations, corresponding to ~20 minutes for one realization.

### G. The effect of dissipation

Sound propagating in air suffers from inevitable dissipation even in the absence of any scattering. In the presence of loss, the S-matrix is not strictly unitary, which may compromise the foundation of our scheme. Here, we show that, in our setup, the loss mainly reduces the singular values of the S-matrix, and it does not affect the effectiveness of our scheme.

In our setup, the loss coefficient of the empty waveguide obtained from fitting the experimental spectra is $\delta = 0.0032\omega_0 \ll \omega_0$, meaning that the loss is weak in our system. Such a small amount of loss produces a complex frequency shift of the form $\boldsymbol{S}(\omega) = \boldsymbol{S}\left(\omega_0 + \frac{\mathrm{i}\delta}{2}\right) \approx \boldsymbol{S}(\omega_0)\left[\boldsymbol{I} - \frac{\delta}{2}\boldsymbol{Q}_\omega(\omega_0)\right]$, where $\boldsymbol{Q}_\omega = -\mathrm{i}\boldsymbol{S}^{-1}\frac{\partial \boldsymbol{S}}{\partial \omega}$ is the Wigner-Smith (WS) time-delay operator[51]. In SI section S5, we show that, under the weak-loss conditions, the singular vectors of $\boldsymbol{S}(\omega)$ remain invariant, whereas the singular values are reduced by a real factor $s'_\mu = \left(1 - \frac{\delta}{2} q_\mu\right)$, with $q_\mu$ being the eigenvalues of $\boldsymbol{Q}_\omega$, namely, the WS time delay. Consequently, normalizing the singular values recovers a lossless S-matrix. Experimentally, this is achieved by scaling each singular value by $\tilde{s}_\mu = s'_\mu / \left|s'_\mu\right|$. The same normalization strategy has also been employed in the characterization of open channels[45]. The validity of this procedure is confirmed by the excellent agreement between the normalized experimental S-matrices and the corresponding lossless numerical simulations (figure 2).

The presence of scatterers in the waveguide may increase the WS time delay, causing an increase in total loss. This increased loss is dependent on the multiple scattering and, therefore, is difficult to handle. Fortunately, such situations can be avoided by selecting configurations with short dwell time from the optimized configurations. To this end, we evaluate the total WS time delay (total dwell time) $\tau = |\mathrm{Tr}\, \boldsymbol{Q}_\omega| = \left|\sum_\mu q_\mu\right|$, which quantifies the overall dissipation. Then, from an ensemble of 40 different optimized realizations (obtained without considering loss), we identify those satisfying $\tau/\langle\tau\rangle \leq 0.9$, where $\langle\tau\rangle$ denotes the average dwell time over 40 realizations. The configurations presented in figure 2(B1, C1, D1) meet this criterion. Complete derivations and additional validation on the effect of dissipation are provided in SI section S5.

**H. Robustness against frequency detuning and positioning errors**

We quantify the similarity of two S-matrices by the Frobenius fidelity $\mathcal{F}(\boldsymbol{A},\boldsymbol{B}) = \frac{1}{2N}\mathrm{tr}\left(\boldsymbol{A}^\dagger\boldsymbol{B}\right)$, where $N$ is the number of channels. Its modulus $|\mathcal{F}| \in [0,1]$ measures amplitude correlation and its argument capture the global phase difference; two unitary matrices differing only by a global phase are regarded as equivalent. For a small frequency detuning $\Delta\omega$, $\boldsymbol{S}(\omega_0+\Delta\omega) \approx \boldsymbol{S}(\omega_0)e^{\mathrm{i}\Delta\omega\boldsymbol{Q}_\omega}$, and the fidelity decays quadratically,

$$|\mathcal{F}| \approx 1 - \frac{1}{4N}(\Delta\omega)^2\left[\mathrm{tr}\,\boldsymbol{Q}_\omega^2 - \frac{(\mathrm{tr}\,\boldsymbol{Q}_\omega)^2}{2N}\right]. \tag{17}$$

Equation (17) has no first-order term and the leading term is quadratic in $\Delta\omega$, which means the S-matrix is insensitive to small changes in frequency. Numerically, for the configuration $\sigma_1$. $|\mathcal{F}|$ between $\boldsymbol{S}(\omega_0+\Delta\omega)$ and $\boldsymbol{S}(\omega_0)$ remains above 90% over 160 Hz. The quadratic scaling ensures that the single-frequency optimization robust against frequency errors (see SI section S6 for the data).

Positioning errors behave analogously: a displacement $\Delta x$ of one scatterer yields $\boldsymbol{S}(x_0+\Delta x) \approx \boldsymbol{S}(x_0)e^{\mathrm{i}\Delta x\boldsymbol{Q}_x}$ and a quadratic fidelity decay of the same form

$$|\mathcal{F}| \approx 1 - \frac{1}{4N}(\Delta x)^2\left[\mathrm{tr}\,\boldsymbol{Q}_x^2 - \frac{(\mathrm{tr}\,\boldsymbol{Q}_x)^2}{2N}\right]. \tag{18}$$

We simulated the manual placement condition (0~0.5 mm error, ~5% of the cylinder radius) with 1000 random realizations. The fidelity $|\mathcal{F}|$ exceeds 0.98 (mean 99.3 %). All cylinders in our experiment were placed manually using coordinates printed on an A4 paper template, confirming that the method is, to a reasonable degree, robust against practical alignment errors (see SI section S7 for the data).

Supplementary information for

# Tailoring Reflectionless Complex Media for non-Abelian Braiding of Acoustic Modes

Hongkuan Zhang[1], Guancong Ma[1†]

[1]Department of Physics, Hong Kong Baptist University, Kowloon Tong, Hong Kong, China

[†]Email: phgcma@hkbu.edu.hk

## S1 Experimental determination of the scattering matrix

Here, we detail the experimental reconstruction of the scattering matrix (S-matrix). A schematic of the setup is shown in figure S1. This waveguide has a width of $W = 0.2$ m, and is operated at $f = 3.2$ kHz, supporting $N = 4$ propagating modes. Let $a_n$ and $d_n$ denote the complex amplitudes of waves incident from the left and right sides of the scattering region, respectively, and $b_n$ and $c_n$ the corresponding outgoing amplitudes. To resolve all propagating modes, the acoustic field is sampled by $4N = 16$ microphones arranged in four groups at longitudinal positions $x_1$, $x_2$, $x_3$, $x_4$. Each group contains $N = 4$ microphones positioned at transverse locations $y_p$ ($p = 1, 2, 3, 4$), which are symmetrically distributed according to $y_p = \frac{W}{N} p - \frac{W}{2N}$.

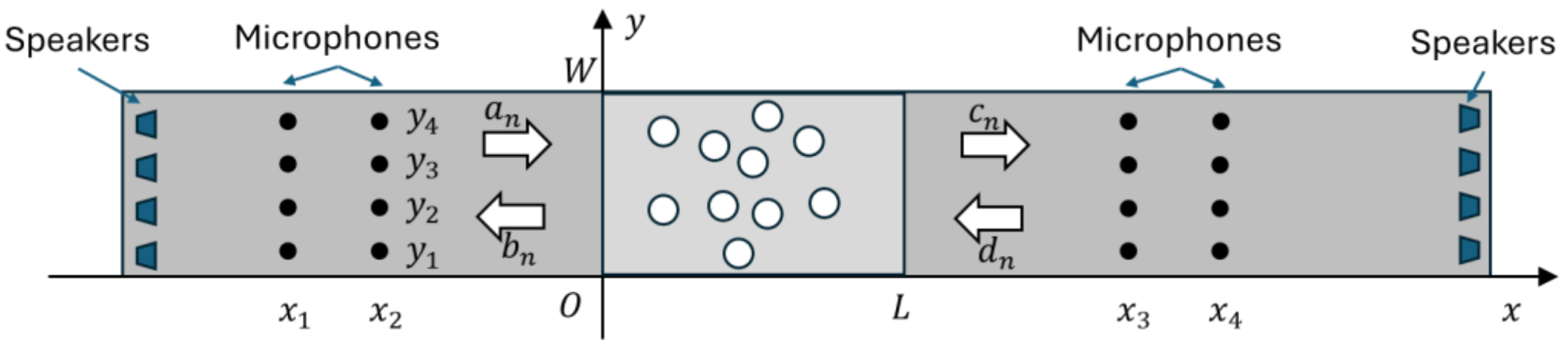

**Figure S1. Schematic diagram of the experimental setup**.

A total of eight loudspeakers are installed in the setup, with $N = 4$ on each side of the waveguide. To construct the S-matrix, the system is excited by each loudspeaker sequentially, and the corresponding acoustic field is recorded by the 16 microphones. This procedure is repeated for all eight speakers. As an example, when only the first speaker (located in the lower left corner) is driven, a scattered pressure field $P(x_j, y_p)$ is generated and measured by the $4N = 16$ microphones, where $j = 1, 2, 3, 4$ denote the longitudinal positions and $p = 1, 2, 3, 4$ denote the transverse positions. The measured pressure on the left side of the sample ($x = x_1, x_2$) is decomposed into waveguide modes as

$$P(x_j, y_p) = \sum_{n=1}^{N} \left( a_n \frac{e^{-\mathrm{i}k_n^x x_j}}{\sqrt{k_n^x}} + b_n \frac{e^{\mathrm{i}k_n^x x_j}}{\sqrt{k_n^x}} \right) \Psi_n(y_p), j = 1,2. \quad \text{(S1.)}$$

and on the right side ($x = x_3, x_4$) as

$$P(x_j, y_p) = \sum_{n=1}^{N} \left( c_n \frac{e^{-\mathrm{i}k_n^x (x_j - L)}}{\sqrt{k_n^x}} + d_n \frac{e^{\mathrm{i}k_n^x (x_j - L)}}{\sqrt{k_n^x}} \right) \Psi_n(y_p), j = 3,4. \quad \text{(S2.)}$$

Here, $n = 1, 2, 3, 4$ denotes the mode index. The transverse mode profile is discretized as

$$\Psi_n(y_p) = \sqrt{\frac{2-\delta_{1n}}{W}}\cos k_n^y y_p\,, \quad \text{(S3.)}$$

$\delta_{mn}$ is the Kronecker delta, with the transverse and axial wavenumbers are given by $k_n^y = (n-1)\pi/W$ and $k_n^x = \sqrt{\omega^2/c_0^2 - (k_n^y)^2}$, respectively ($\omega = 2\pi f$, $c_0$ is the sound speed in air, and $L$ is the length of the scattering region). These modes satisfy the discrete orthogonality condition[1]

$$\sum_{p=1}^{N} \Psi_m(y_p)\Psi_n(y_p)\frac{W}{N} = \delta_{mn}. \quad \text{(S4.)}$$

This relationship can be used to decouple the coefficients in Eq. (S1-S2). For example, projecting Eq. (S1) onto the modal basis gives

$$\sum_{p=1}^{N} P(x_j, y_p)\Psi_m(y_p)\frac{W}{N} = \sum_{p=1}^{N}\sum_{n=1}^{N}\left(a_n\frac{e^{-\mathrm{i}k_n^x x_j}}{\sqrt{k_n^x}} + b_n\frac{e^{\mathrm{i}k_n^x x_j}}{\sqrt{k_n^x}}\right)\Psi_n(y_p)\Psi_m(y_p)\frac{W}{N}, j = 1,2. \quad \text{(S5.)}$$

Using the orthogonality relation Eq. (S4), this simplifies to

$$\sqrt{k_n^x}\frac{W}{N}\sum_{p=1}^{N} P(x_j, y_p)\Psi_n(y_p) = a_n e^{-\mathrm{i}k_n^x x_j} + b_n e^{\mathrm{i}k_n^x x_j}, j = 1,2. \quad \text{(S6.)}$$

Similar results can be obtained for points $x_3$ and $x_4$ as well. Therefore, the incident and outgoing coefficients for each waveguide mode are given by

$$\begin{cases} a_n = \dfrac{W}{N}\dfrac{\sqrt{k_n^x}}{2\mathrm{i}\sin[k_n^x(x_2-x_1)]}\sum_{p=1}^{N}\left[-P(x_2,y_p)e^{\mathrm{i}k_n^x x_1} + P(x_1,y_p)e^{\mathrm{i}k_n^x x_2}\right]\Psi_n(y_p), \\ b_n = \dfrac{W}{N}\dfrac{\sqrt{k_n^x}}{2\mathrm{i}\sin[k_n^x(x_2-x_1)]}\sum_{p=1}^{N}\left[P(x_2,y_p)e^{-\mathrm{i}k_n^x x_1} - P(x_1,y_p)\,e^{-\mathrm{i}k_n^x x_2}\right]\Psi_n(y_p), \\ c_n = \dfrac{W}{N}\dfrac{\sqrt{k_n^x}}{2\mathrm{i}\sin[k_n^x(x_4-x_3)]}\sum_{p=1}^{N}\left[-P(x_4,y_p)e^{\mathrm{i}k_n^x(x_3-L)} + P(x_3,y_p)e^{\mathrm{i}k_n^x(x_4-L)}\right]\Psi_n(y_p), \\ d_n = \dfrac{W}{N}\dfrac{\sqrt{k_n^x}}{2\mathrm{i}\sin[k_n^x(x_4-x_3)]}\sum_{p=1}^{N}\left[P(x_4,y_p)e^{-\mathrm{i}k_n^x(x_3-L)} - P(x_3,y_p)e^{-\mathrm{i}k_n^x(x_4-L)}\right]\Psi_n(y_p), \end{cases} \quad \text{(S7.)}$$

for $n = 1,2,3,4$. Here, we directly converted the data from the 16 microphones into 16 coefficients.

The S-matrix, which relates the vectors of incidence $|\psi_{\mathrm{in}}\rangle = (a_1, \dots, a_N, d_1, \dots, d_N)^{\mathrm{T}}$ to outgoing $|\psi_{\mathrm{out}}\rangle = (b_1, \dots, b_N, c_1, \dots, c_N)^{\mathrm{T}}$ via $|\psi_{\mathrm{out}}\rangle = \boldsymbol{S}|\psi_{\mathrm{in}}\rangle$, has dimension of $2N \times 2N$ (here $8 \times 8$). Each independent excitation of the $2N = 8$ loudspeakers provides one column of the relation

$$\begin{bmatrix} b_1^{(1)} & b_1^{(2)} & \cdots & b_1^{(2N)} \\ \vdots & \vdots & \vdots & \vdots \\ b_N^{(1)} & b_N^{(2)} & \dots & b_N^{(2N)} \\ c_1^{(1)} & c_1^{(2)} & \dots & c_1^{(2N)} \\ \vdots & \vdots & \vdots & \vdots \\ c_N^{(1)} & c_N^{(2)} & \cdots & c_N^{(2N)} \end{bmatrix} = \boldsymbol{S} \begin{bmatrix} a_1^{(1)} & a_1^{(2)} & \cdots & a_1^{(2N)} \\ \vdots & \vdots & \vdots & \vdots \\ a_N^{(1)} & a_N^{(2)} & \dots & a_N^{(2N)} \\ d_1^{(1)} & d_1^{(2)} & \dots & d_1^{(2N)} \\ \vdots & \vdots & \vdots & \vdots \\ d_N^{(1)} & d_N^{(2)} & \cdots & d_N^{(2N)} \end{bmatrix}, \tag{S8.}$$

where superscripts denote the loudspeaker indices and subscripts indicate mode numbers computed via Eq. (S7). Consequently, exciting all loudspeaker sequentially supplies all the needed data. The S-matrix is then

$$\boldsymbol{S} = \begin{bmatrix} b_1^{(1)} & b_1^{(2)} & \cdots & b_1^{(2N)} \\ \vdots & \vdots & \vdots & \vdots \\ b_N^{(1)} & b_N^{(2)} & \dots & b_N^{(2N)} \\ c_1^{(1)} & c_1^{(2)} & \dots & c_1^{(2N)} \\ \vdots & \vdots & \vdots & \vdots \\ c_N^{(1)} & c_N^{(2)} & \cdots & c_N^{(2N)} \end{bmatrix} \begin{bmatrix} a_1^{(1)} & a_1^{(2)} & \cdots & a_1^{(2N)} \\ \vdots & \vdots & \vdots & \vdots \\ a_N^{(1)} & a_N^{(2)} & \dots & a_N^{(2N)} \\ d_1^{(1)} & d_1^{(2)} & \dots & d_1^{(2N)} \\ \vdots & \vdots & \vdots & \vdots \\ d_N^{(1)} & d_N^{(2)} & \cdots & d_N^{(2N)} \end{bmatrix}^{-1}. \tag{S9.}$$

In principle, more than $2N$ measurements can be used to enhance reconstruction accuracy. In such cases, the matrix inversion in Eq. (S9) is replaced by the Moore-Penrose pseudo-inverse.

## S2 Validation of the Yang-Baxter equation

One fascinating characteristic of the braid generators is given by $\sigma_i \sigma_j \sigma_i = \sigma_j \sigma_i \sigma_j$ for $|i - j| = 1$, which is also known as the is the Yang-Baxter equation. In fact, one way to define a braid group is through this relation, as displayed in Eq. (3) in the main text. Here, we present a numerical validation of this relation in our multiple scattering settings. Figures S2(A, B) show the braiding diagrams and the corresponding T-matrices of three scattering sections. In both cases, modes $|1\rangle, |2\rangle, |3\rangle$ are swapped and mode $|4\rangle$ is unaffected. Figures S2(C, D) display the simulated field patterns, confirming the expected behavior. In both cases, incident waves in channels $|1\rangle$, $|2\rangle$, $|3\rangle$, and $|4\rangle$ are converted to $e^{-\mathrm{i}\pi/3}|3\rangle$, $-|2\rangle$, $e^{\mathrm{i}\pi/3}|1\rangle$, and $|4\rangle$ at the output.

Experimental validation is currently beyond our capability, owing to the space limitation of our lab.

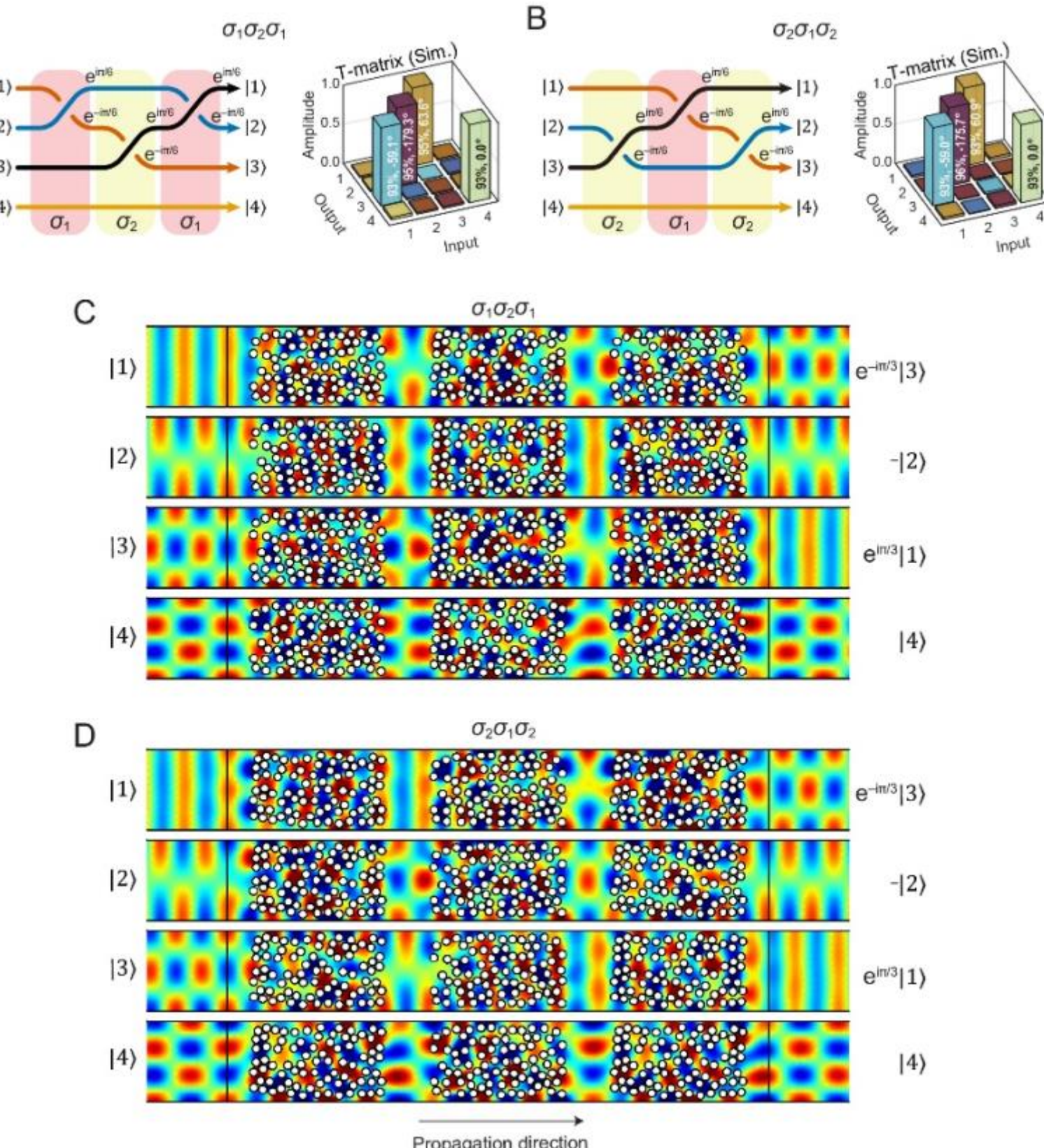

**Figure S2.** Demonstration of the Yang-Baxter relation. (A, B) Braiding diagram (left) and simulated T-matrix (right) of the sequences $\sigma_1\sigma_2\sigma_1$ (A) and $\sigma_2\sigma_1\sigma_2$ (B), which are clearly equivalent. (C, D) The simulated acoustic fields of three concatenated scattering sections realizing $\sigma_1\sigma_2\sigma_1$ (C) and $\sigma_2\sigma_1\sigma_2$ (D) The two sequences are equivalent.

## S3 Realizing discrete Fourier transform operations

The discrete Fourier transform (DFT) of a $N \times 1$ vector $\boldsymbol{x} = (x_0, x_1, \cdots, x_{N-1})^{\mathrm{T}}$ is given by $\boldsymbol{X} = \sqrt{N} F_N \boldsymbol{x}$, where $F_N$ is an $N \times N$ unitary matrix with entries $(F_N)_{\mu\nu} = \omega^{(\mu-1)(\nu-1)}$ for $\mu, \nu = 1, \cdots, N$. Here, $\omega = e^{-2\pi \mathrm{i}/N}$ is the primitive $N$-th root of unity. In full matrix form,

$$F_N = \frac{1}{\sqrt{N}} \begin{pmatrix} 1 & 1 & 1 & \cdots & 1 \\ 1 & \omega & \omega^2 & \cdots & \omega^{(N-1)} \\ 1 & \omega^2 & \omega^4 & \cdots & \omega^{2(N-1)} \\ \vdots & \vdots & \vdots & \ddots & \vdots \\ 1 & \omega^{(N-1)} & \omega^{2(N-1)} & \cdots & \omega^{(N-1)(N-1)} \end{pmatrix}. \tag{S10.}$$

Take $N = 8$ as an example, we simulate the DFT by sampling the discrete sequence $x_n = \sin(2\pi f_0 n/N)$ for $n = (0,1, \dots ,7)$. Its DFT spectrum $X_k$ $(k = 0,1, \dots ,7)$ is non-zero only at $k = f_0$ and $k = N - f_0$, which corresponding to the positive and negative spectrum components, respectively.

Because $F_8$ is unitary, the operation can be realized with an $8 \times 8$ T-matrix realized by a suitably designed complex medium embedded in an 8-channel waveguide system. We employ 200 cylindrical scatterers occupying $0.4 \times 0.7\ \mathrm{m}^2$ in an acoustic waveguide 0.4 m in width (twice the width of the four-mode waveguide described in the main text) to achieve $F_8$ at 3.2 kHz. The optimized scatterer configuration is shown in figure S3(B).

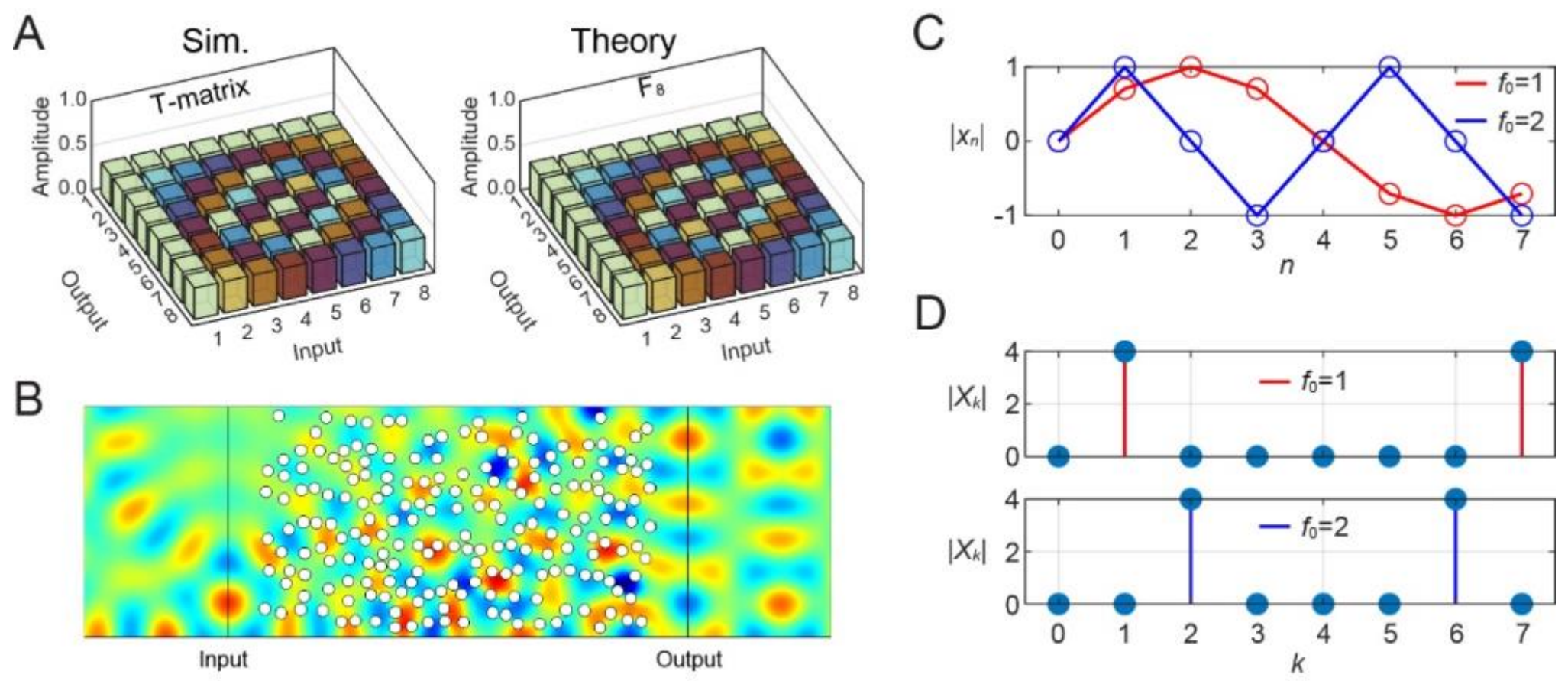

**Figure S3.** Implementation of DFT operations. (A) Left: The T-matrix of the optimized medium implementing the eight channel DFT operation, obtained by full-wave simulation using a suitably designed complex medium. Right: an ideal $8 \times 8$ matrix for the DFT operation. (B-D) Verification of the DFT operation using 8-point sequence of $f_0 = 1$ and $f_0 = 2$. (B) The simulated field pattern obtained by injecting an 8-point signal encoded as modal coefficients ($f_0 = 1$). (C) The input modal coefficients representing the discrete sequence $\{x_n\}$ for the two test cases. (D) The output modal coefficients (i.e., the computed spectrum) extracted from the wave propagated through the complex medium, showing the expected spectral peaks at the corresponding indices.

We demonstrate the DFT operation for two sequences with $f_0 = 1$ and $f_0 = 2$, representing one and two periods per 8 sample points, respectively. The 8 elements of each sequence $\{x_n\}$ are assigned sequentially to the 8 input channels, forming input vector as $\psi_{\text{in}} = \sum_{n=0}^{7} x_n \Psi_{n+1}$ [figure S3(C)]. The output vector is given by $\psi_{\text{out}} = \sum_{k=0}^{7} X_k \Psi_{k+1}$, where $X_k$ denotes the coefficient associated with $(k+1)$-th output channel $\Psi_{k+1}$ [figure S3(D), top and bottom]. The simulated field pattern for the case $f_0 = 1$ under left-side excitation by $\psi_{\text{in}}$ is also shown in figure S3(B). Note that because the T-matrix is in a channel basis, we need to extract the result from the field patterns using the channel basis. As expected, the spectrum for the sequence with $f_0 = 1$, the spectrum $X_k$ exhibits non-zero components only at $k = 1$ (positive-spectrum) and $k = 7$ (negative-spectrum) indices. Similarly, the sequence with $f_0 = 2$ produces non-zero components only at $k = 2$ and $k = 6$.

## S4 The generalized Wigner-Smith operator in acoustics

The generalized Wigner-Smith operator (GWSO) has been studied in contexts such as microwaves and photonics [2–4], and work has extended it to acoustic systems [5,6]. Here, we use a complementary shape-derivative formulation at the operator level to construct the GWSO for a multimode acoustic waveguide. While equivalent results can be obtained from the variational framework of ref. [5], the present derivation via the Gateaux derivative offers an alternative starting point that directly links the GWSO to boundary-integral expressions convenient for numerical implementation.

We first outline the operator-based approach that leads to an integral representation of the GWSO, a framework extensively developed and validated in microwave systems. For comparison, we also recall the expression for the acoustic radiation force, which reveals the proportionality between the force and the GWSO [5].

### 4.1 Expectation value of the GWSO

The following derivation generalizes the approach of Supplemental Ref. [3] to a broader class of wave systems. Following the established formalism, the expectation value of the GWSO (associated with a parameter $\alpha$) can be written as a derivative of the wave operator

$$\langle \psi_\mu | \boldsymbol{Q}_\alpha | \psi_\nu \rangle = -\frac{1}{2} \left\langle P_{\psi_\mu} \middle| \frac{\partial \mathcal{L}}{\partial \alpha} \middle| P_{\psi_\nu} \right\rangle, \tag{S11.}$$

where $|\psi_\mu\rangle$ is the $\mu$-th incident waveguide mode and $|P_{\psi_\mu}\rangle$ is the corresponding acoustic pressure field inside the scattering region. For acoustic waves,

$$\mathcal{L} = \rho(\boldsymbol{r})[\boldsymbol{\nabla} \cdot (\rho^{-1}(\boldsymbol{r}) \boldsymbol{\nabla}) + \kappa^{-1}(\boldsymbol{r}) \omega^2], \tag{S12.}$$

where $\kappa(\boldsymbol{r})$ and $\rho(\boldsymbol{r})$ are the position-dependent bulk modulus and mass density and $\omega = 2\pi f$ is the angular frequency. Our goal is to evaluate the operator derivative $\frac{\partial \mathcal{L}}{\partial \alpha}$. To describe variations induced by a geometric parameter $\alpha$, we adopt the Gateaux derivative method[7–9], also referred to as the velocity method or domain-deformation method.

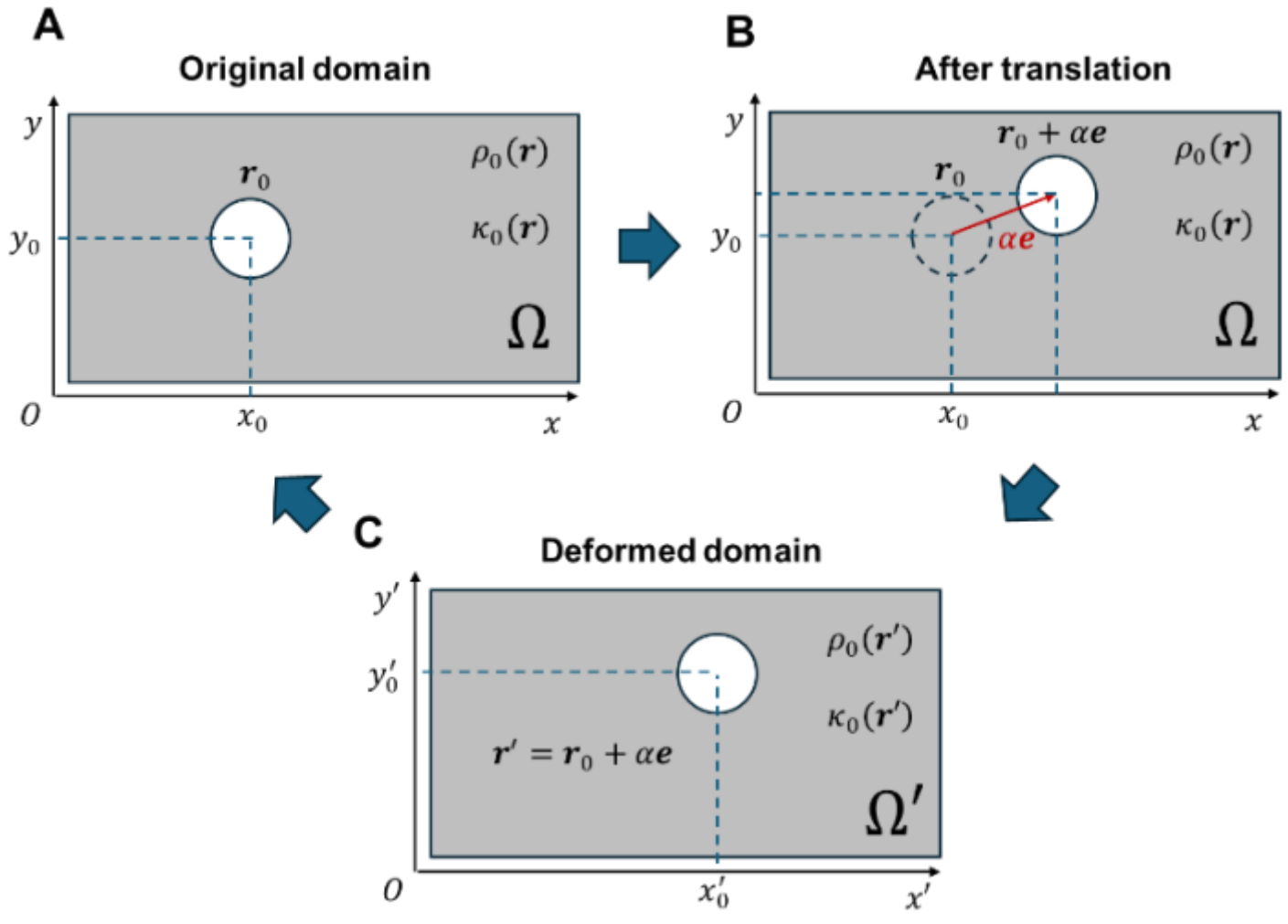

**Figure S4.** Schematic diagram of Gateaux derivative method.

We focus on the case where $\alpha$ represents a translation of a scatterer. The idea is illustrated in figure S4. Consider a scatterer initially located at $\boldsymbol{r}_0$ inside a region $\Omega$ [figure S4(A)]. The scatterer is displaced by a small amount $\alpha$ along a unit direction $\boldsymbol{e}$, so that its new position is $\boldsymbol{r}' = \boldsymbol{r}_0 + \alpha\boldsymbol{e}$ [figure S4(B)]. The change in the scatterer position results in a change in the domain $\Omega$, which is the governing region of the acoustic wave equation, or equivalently the linear operator $\mathcal{L}$ [Eq. (S12)]. The change in domain thus causes a change in the operator, which must be computed the accurately predict the change in sound field. The Gateaux-derivative framework treats the motion as a continuous mapping of parameter distribution [figure S4(C)]. Under this viewpoint, the translation is described by the mapping

$$\mathcal{T}_\alpha(\boldsymbol{r}) = \boldsymbol{r} + \boldsymbol{v}(\boldsymbol{r}), \tag{S13.}$$

where $\boldsymbol{v}(\boldsymbol{r})$ is a "virtual velocity" that deforms the original domain $\Omega$ into a new domain $\Omega'$. For small $\alpha$, the deformation gradient, its inverse, and the Jacobian determinant are approximated

$$\boldsymbol{F} = \mathcal{T}_\alpha(\boldsymbol{r})\boldsymbol{\nabla} = \boldsymbol{I} + \alpha\boldsymbol{v}(\boldsymbol{r})\boldsymbol{\nabla}, \tag{S14.}$$

$$\boldsymbol{F}^{-1} = \boldsymbol{I} - \alpha\boldsymbol{v}(\boldsymbol{r})\boldsymbol{\nabla} + O(\alpha^2), \tag{S15.}$$

$$J = \det\boldsymbol{F} = 1 + \alpha\boldsymbol{\nabla}\cdot\boldsymbol{v}(\boldsymbol{r}) + O(\alpha^2), \tag{S16.}$$

where $\boldsymbol{\nabla}$ denotes the gradient operator in the initial domain $\Omega$. Quantities in the deformed domain $\Omega'$ will be marked with a prime. Note that, $\boldsymbol{F}$ is a two-point tensor, with its bases defined in different spaces. The operator derivative is defined as

$$\frac{\partial \mathcal{L}_\Omega}{\partial \alpha} = \lim_{\alpha\to 0^+} \frac{\mathcal{L}_{\Omega'} - \mathcal{L}_\Omega}{\alpha} = \frac{\mathrm{d}}{\mathrm{d}\alpha}(\mathcal{L}_{\Omega'} \circ \mathcal{T}_\alpha)\Big|_{\alpha=0}, \tag{S17.}$$

where $\circ\, \mathcal{T}_\alpha$ is a pullback to the original domain $\Omega$. The reason for this pullback mapping is that our operator derivative should be defined in the initial domain (coordinates). In the deformed domain, the wave operator reads

$$\mathcal{L}_{\Omega'} = \rho(\boldsymbol{r}')[\boldsymbol{\nabla}' \cdot (\rho^{-1}(\boldsymbol{r}')\boldsymbol{\nabla}') + \omega^2\kappa^{-1}(\boldsymbol{r}')]. \tag{S18.}$$

Pulling it back to $\Omega$ gives

$$\mathcal{L}_{\Omega'} \circ \mathcal{T}_\alpha = \rho\big(\boldsymbol{r} + \alpha\boldsymbol{v}(\boldsymbol{r})\big)\big[J^{-1}\boldsymbol{\nabla} \cdot \big(J\boldsymbol{F}^{-1}\rho^{-1}\big(\boldsymbol{r} + \alpha\boldsymbol{v}(\boldsymbol{r})\big)\boldsymbol{F}^{-\mathrm{T}}\boldsymbol{\nabla}\big) + \omega^2\kappa^{-1}\big(\boldsymbol{r} + \alpha\boldsymbol{v}(\boldsymbol{r})\big)\big], \tag{S19.}$$

where we used the identities

$$\boldsymbol{\nabla}' \cdot \mathbf{A} = J^{-1}\boldsymbol{\nabla} \cdot (J\boldsymbol{F}^{-1}\mathbf{A}) \text{ and } \boldsymbol{\nabla}' = \boldsymbol{F}^{-\mathrm{T}}\boldsymbol{\nabla}. \tag{S20.}$$

Importantly, the pullback procedure shows that the change in material parameters due to the change in domain can be expressed in the original coordinate, e.g., $\rho(\boldsymbol{r}') \to \rho\big(\boldsymbol{r} + \alpha\boldsymbol{v}(\boldsymbol{r})\big)$. Expanding the material parameters to first order in $\alpha$

$$\rho\big(\boldsymbol{r} + \alpha\boldsymbol{v}(\boldsymbol{r})\big) = \rho(\boldsymbol{r}) + \alpha\boldsymbol{\nabla}\rho(\boldsymbol{r}) \cdot \boldsymbol{v}(\boldsymbol{r}) + O(\alpha^2). \tag{S21.}$$

and similarly, for $\rho^{-1}\big(\boldsymbol{r} + \alpha\boldsymbol{v}(\boldsymbol{r})\big)$ and $\kappa^{-1}\big(\boldsymbol{r} + \alpha\boldsymbol{v}(\boldsymbol{r})\big)$. Substituting these expansions into Eq. (S19), then we obtain

$$\begin{aligned}\mathcal{L}_{\Omega'} \circ \mathcal{T}_\alpha \approx \frac{(\rho + \alpha\boldsymbol{\nabla}\rho\cdot\boldsymbol{v})}{(1 + \alpha\boldsymbol{\nabla}\cdot\boldsymbol{v})}\boldsymbol{\nabla} \cdot [(1 + \alpha\boldsymbol{\nabla}\cdot\boldsymbol{v})(\boldsymbol{I} - \alpha\boldsymbol{v}\boldsymbol{\nabla})(\rho^{-1} + \alpha\boldsymbol{\nabla}(\rho^{-1})\cdot\boldsymbol{v})(\boldsymbol{I} - \alpha\boldsymbol{v}\boldsymbol{\nabla})^{\mathrm{T}}\boldsymbol{\nabla}]\\ +\omega^2(\rho + \alpha\boldsymbol{\nabla}\rho\cdot\boldsymbol{v})(\kappa^{-1} + \alpha\boldsymbol{\nabla}(\kappa^{-1})\cdot\boldsymbol{v}).\end{aligned} \tag{S22.}$$

Using Eq. (S17), and differentiating Eq. (S22) with respect to $\alpha$, and evaluating at $\alpha = 0$ yields

$$\begin{aligned}\frac{\partial \mathcal{L}_\Omega}{\partial \alpha} = \rho\boldsymbol{\nabla} \cdot [(\boldsymbol{\nabla}(\rho^{-1})\cdot\boldsymbol{v})\boldsymbol{\nabla}] + \rho\boldsymbol{\nabla} \cdot [(-\boldsymbol{v}\boldsymbol{\nabla} - \boldsymbol{\nabla}\boldsymbol{v})\rho^{-1}\boldsymbol{\nabla}] + \rho\boldsymbol{\nabla} \cdot (\rho^{-1}(\boldsymbol{\nabla}\cdot\boldsymbol{v})\boldsymbol{\nabla})\\ -\rho(\boldsymbol{\nabla}\cdot\boldsymbol{v})\boldsymbol{\nabla} \cdot \big((\rho^{-1})\boldsymbol{\nabla}\big) + \omega^2\rho(\boldsymbol{\nabla}(\kappa^{-1})\cdot\boldsymbol{v}) + (\boldsymbol{\nabla}\rho\cdot\boldsymbol{v})[\boldsymbol{\nabla}\cdot(\rho^{-1}\boldsymbol{\nabla}) + \omega^2\kappa^{-1}].\end{aligned} \tag{S23.}$$

For a translation, the velocity field $\boldsymbol{v}$ is constant ($\boldsymbol{v} = \boldsymbol{e}$), so that $\boldsymbol{\nabla}\boldsymbol{v} + \boldsymbol{v}\boldsymbol{\nabla} = \boldsymbol{0}$ (zero strain rate) and $\boldsymbol{\nabla}\cdot\boldsymbol{v} = \boldsymbol{0}$ (zero volume strain rate). Moreover, the last term in Eq. (S23) vanishes because the pressure field satisfies the homogeneous Helmholtz equation $\boldsymbol{\nabla}\cdot(\rho^{-1}\boldsymbol{\nabla}P) + \omega^2\kappa^{-1}P = 0$. Therefore, the operator derivative simplifies to

$$\frac{\partial \mathcal{L}_\Omega}{\partial \alpha} = \rho \boldsymbol{\nabla} \cdot [(\boldsymbol{\nabla}(\rho^{-1}) \cdot \boldsymbol{e})\boldsymbol{\nabla}] + \rho\omega^2(\boldsymbol{\nabla}(\kappa^{-1}) \cdot \boldsymbol{e}). \tag{S24.}$$

For scatterers with piecewise constant material parameters, the coefficients can be expressed as step functions $H(\boldsymbol{r} - \boldsymbol{r}_\Gamma)$ across the interface $\Gamma$

$$\rho^{-1}(\boldsymbol{r}) = \rho_0^{-1} + (\rho_1^{-1} - \rho_0^{-1})H(\boldsymbol{r} - \boldsymbol{r}_\Gamma). \tag{S25.}$$

Here, the subscript 1 represents the acoustic parameters of the scatterer, and 0 represents the acoustic parameters of the background medium (air). Thus, the gradient is

$$\boldsymbol{\nabla}(\rho^{-1}) = (\rho_1^{-1} - \rho_0^{-1})\delta(\boldsymbol{r} - \boldsymbol{r}_\Gamma)\boldsymbol{n}, \tag{S26.}$$

where $\delta(\boldsymbol{r} - \boldsymbol{r}_\Gamma)$ is the Dirac delta supported on $\Gamma$, and $\boldsymbol{n} = (n_x, n_y)$ is the outward unit normal. We consider the scatterers to be sufficiently massive and rigid, with density and bulk modulus approaching infinity compared to air, so

$$(\rho_1^{-1} - \rho_0^{-1}) = \lim_{\rho_1 \to \infty}\left(\frac{1}{\rho_1} - \frac{1}{\rho_0}\right) = -\frac{1}{\rho_0}. \tag{S27.}$$

$$(\kappa_1^{-1} - \kappa_0^{-1}) = \lim_{\kappa_1 \to \infty}\left(\frac{1}{\kappa_1} - \frac{1}{\kappa_0}\right) = -\frac{1}{\kappa_0} = -\frac{1}{\rho_0 c_0^2}. \tag{S28.}$$

In addition, we are examining the operator derivative in the background region. The external coefficient $\rho$ in Eq. (S24) should take a value outside the scatterer, i.e., $\rho_0$. Thus, the operator derivative Eq. (S24) becomes

$$\frac{\partial \mathcal{L}_\Omega}{\partial \alpha} = -\boldsymbol{\nabla} \cdot [\delta(\boldsymbol{r} - \boldsymbol{r}_\Gamma)(\boldsymbol{n} \cdot \boldsymbol{e})\boldsymbol{\nabla}] - k_0^2 \delta(\boldsymbol{r} - \boldsymbol{r}_\Gamma)(\boldsymbol{n} \cdot \boldsymbol{e}), \tag{S29.}$$

with $k_0 = \omega / c_0$. Inserting Eq. (S29) into Eq. (S11) yields

$$\langle \psi_\mu | \boldsymbol{Q}_\alpha | \psi_\nu \rangle = \frac{1}{2}\int_\Omega \left\{ P_{\psi_\mu}^* \boldsymbol{\nabla} \cdot \left[\delta(\boldsymbol{r} - \boldsymbol{r}_\Gamma)(\boldsymbol{n} \cdot \boldsymbol{e})\boldsymbol{\nabla} P_{\psi_\nu}\right] + k_0^2 \delta(\boldsymbol{r} - \boldsymbol{r}_\Gamma)(\boldsymbol{n} \cdot \boldsymbol{e}) P_{\psi_\mu}^* P_{\psi_\nu} \right\} \mathrm{d}\Omega. \tag{S30.}$$

Applying the divergence theorem and using continuity of acoustic pressure and normal velocity across the interface, the first term reduces to a tangential-gradient contribution. The final result for a rigid boundary is

$$\langle \psi_\mu | \boldsymbol{Q}_\alpha | \psi_\nu \rangle = \frac{1}{2}\int_\Gamma k_0^2 P_{\psi_\mu}^* P_{\psi_\nu} (\boldsymbol{n} \cdot \boldsymbol{e}) \mathrm{d}\Gamma - \frac{1}{2}\int_\Gamma \partial_\parallel P_{\psi_\mu}^* \, \partial_\parallel P_{\psi_\nu} (\boldsymbol{n} \cdot \boldsymbol{e}) \mathrm{d}\Gamma. \tag{S31.}$$

where $\partial_\parallel = \boldsymbol{\nabla} - \boldsymbol{n}(\boldsymbol{n} \cdot \boldsymbol{\nabla})$ denotes the tangential gradient along the boundary $\Gamma$. Thus, the GWSO for the translation of the $m$-th scatterer, it is sufficient to perform the boundary integral over its surface

$$\begin{pmatrix} Q_{x_m}^{\mu\nu} \\ Q_{y_m}^{\mu\nu} \end{pmatrix} = \frac{1}{2}\int_{\Gamma_m} \left( k_0^2 P_{\psi_\mu}^* P_{\psi_\nu} - \partial_\parallel P_{\psi_\mu}^* \partial_\parallel P_{\psi_\nu} \right) \begin{pmatrix} n_x \\ n_y \end{pmatrix} \mathrm{d}\Gamma_m \,. \tag{S32.}$$

For the sake of completeness, we also consider scatterers that are treated as sound-soft boundary, i.e., the pressure vanishes at the interface. A direct limit of vanishing density and bulk modulus in the material-parameter formulation Eq. (S27-S28) leads to a divergent result. Therefore, the derivation must be adapted to avoid the singularity. Starting from Eq. (S24) in the sound-hard boundary derivation, the operator derivative for a translation along a unit vector $\boldsymbol{e}$ can also be written as

$$\frac{\partial \mathcal{L}_\Omega}{\partial \alpha} = \rho_0 \boldsymbol{\nabla} \cdot \left[ \left( -\frac{\boldsymbol{\nabla}\rho}{\rho_0^2} \cdot \boldsymbol{e} \right) \boldsymbol{\nabla} \right] + \rho_0 \omega^2 \left( -\frac{\boldsymbol{\nabla}\kappa}{\kappa_0^2} \cdot \boldsymbol{e} \right). \tag{S33.}$$

For a soft scatterer, both the density and the bulk modulus approach zero inside the object, i.e., $\rho_1 = 0, \kappa_1 = 0$. Hence the gradients of the material parameters become concentrated on the interface $\Gamma$

$$\boldsymbol{\nabla}\rho = -\rho_0 \delta(\boldsymbol{r} - \boldsymbol{r}_\Gamma)\boldsymbol{n}, \tag{S34.}$$

$$\boldsymbol{\nabla}\kappa = -\kappa_0 \delta(\boldsymbol{r} - \boldsymbol{r}_\Gamma)\boldsymbol{n}. \tag{S35.}$$

Substituting them into Eq. (S33) yields

$$\frac{\partial \mathcal{L}_\Omega}{\partial \alpha} = \boldsymbol{\nabla} \cdot [(\delta(\boldsymbol{r} - \boldsymbol{r}_\Gamma)\boldsymbol{n} \cdot \boldsymbol{e})\boldsymbol{\nabla}] + k_0^2 (\delta(\boldsymbol{r} - \boldsymbol{r}_\Gamma) \cdot \boldsymbol{e}), \tag{S36.}$$

which differs from the sound-hard result (Eq. (S29)) by a sign. Inserting Eq. (S36) into Eq. (S11) gives

$$\langle \psi_\mu | \boldsymbol{Q}_\alpha | \psi_\nu \rangle = -\frac{1}{2}\int_\Omega \left\{ P_{\psi_\mu}^* \boldsymbol{\nabla} \cdot \left[ \delta(\boldsymbol{r} - \boldsymbol{r}_\Gamma)(\boldsymbol{n} \cdot \boldsymbol{e}) \boldsymbol{\nabla} P_{\psi_\nu} \right] + k_0^2 \delta(\boldsymbol{r} - \boldsymbol{r}_\Gamma)(\boldsymbol{n} \cdot \boldsymbol{e}) P_{\psi_\mu}^* P_{\psi_\nu} \right\} \mathrm{d}\Omega \,. \tag{S37.}$$

Applying the divergence theorem and noting that on a soft boundary the pressure vanishes everywhere and thus its tangential gradient is also zero, the volume integral reduces to a surface integral involving only the normal derivatives

$$\langle \psi_\mu | \boldsymbol{Q}_\alpha | \psi_\nu \rangle = \frac{1}{2}\int_\Gamma \frac{\partial P_{\psi_\mu}^*}{\partial n} \frac{\partial P_{\psi_\nu}}{\partial n} (\boldsymbol{n} \cdot \boldsymbol{e}) \mathrm{d}\Gamma \,. \tag{S38.}$$

For the translation of the $m$-th scatterer, the operator components are

$$\begin{pmatrix} Q_{x_m}^{\mu\nu} \\ Q_{y_m}^{\mu\nu} \end{pmatrix} = \frac{1}{2}\int_{\Gamma_m} \left( \partial_\perp P_{\psi_\mu}^* \partial_\perp P_{\psi_\nu} \right) \begin{pmatrix} n_x \\ n_y \end{pmatrix} \mathrm{d}\Gamma_m \,, \tag{S39.}$$

where $\partial_\perp = \boldsymbol{n} \cdot \boldsymbol{\nabla} = \partial / \partial n$ denotes the normal gradient along the boundary.

This derivation shows that the GWSO can be expressed using boundary integrals when the perturbation corresponds to a translation of the scatterer. For rigid scatterers that can be considered as sound-hard boundaries, the operator derivative involves the tangential gradient, which differs from the electromagnetic case. The result for rigid scatterers coincides with that of ref. [5]. For sound-soft boundaries, the formula involves the normal derivative, which is directly analogous to the electromagnetic GWSO[4].

### 4.2 Acoustic radiation force

The acoustic radiation force can be described by the Brillouin stress tensor[9,10]

$$\boldsymbol{T} = \frac{1}{2}(\beta p^2 - \rho|\boldsymbol{v}|^2)\boldsymbol{I} + \rho\boldsymbol{v} \otimes \boldsymbol{v}. \tag{S40.}$$

Note that all field quantities are understood in the time-averaged sense, i.e., $P \to \langle P \rangle$, though the angle brackets are omitted here for notational simplicity. The total radiation force acting on a particle is given by a surface integral over the boundary of the scatterer $\Gamma$

$$\boldsymbol{F} = -\oint_{\Gamma} \boldsymbol{n} \cdot \boldsymbol{T} \mathrm{d}\Gamma\,. \tag{S41.}$$

If a discontinuity exists across the interface, the expression becomes

$$\boldsymbol{F} = \oint_{\Gamma} \boldsymbol{n} \cdot (\boldsymbol{T}|_{\mathrm{out}} - \boldsymbol{T}|_{\mathrm{in}}) \mathrm{d}\Gamma\,. \tag{S42.}$$

Let the outer medium have parameters $\beta_0$ and $\rho_0$, and the inner material $\beta_1$and $\rho_1$. Denote by $\boldsymbol{n}$ and $\boldsymbol{t}$, the normal ($\perp$) and tangential ($\parallel$) directions, respectively. Using the continuity of acoustic pressure $P|_{\mathrm{out}} = P|_{\mathrm{in}}$, and normal velocity $v_\perp|_{\mathrm{out}} = v_\perp|_{\mathrm{in}}$, and noting that

$$\boldsymbol{n} \cdot \boldsymbol{v} \otimes \boldsymbol{v} = v_\perp(v_\perp\boldsymbol{n} + v_\parallel\boldsymbol{t}) = v_\perp^2\boldsymbol{n} + v_\perp v_\parallel \boldsymbol{t}, \tag{S43.}$$

the radiation force can be written as

$$\boldsymbol{F} = \oint_{\Gamma} \begin{bmatrix} \frac{1}{2}\left((\beta_0 - \beta_1)P^2 - (\rho_0 - \rho_1)v_\perp^2 - \rho_0 v_\parallel^2\big|_{\mathrm{out}} + \rho_1 v_\parallel^2\big|_{\mathrm{in}}\right)\boldsymbol{n} \\ +(\rho_0 - \rho_1)v_\perp^2\boldsymbol{n} + \left(\rho_0 v_\parallel\big|_{\mathrm{out}} - \rho_1 v_\parallel\big|_{\mathrm{in}}\right) v_\perp \boldsymbol{t} \end{bmatrix} \mathrm{d}\Gamma\,. \tag{S44.}$$

For rigid boundaries, where $\beta_1 = 0$, $v_\perp = 0$ and $v_\parallel\big|_{\mathrm{in}} = 0$, the expression simplifies to

$$\boldsymbol{F} = \oint_{\Gamma} \frac{1}{2}\left(\beta_0 P^2 - \rho_0 v_\parallel^2\big|_{\mathrm{out}}\right) \boldsymbol{n} \mathrm{d}\Gamma\,. \tag{S45.}$$

For pressure-release boundaries, where $\rho_1 = 0$, $P = 0$ and $v_\parallel\big|_{\mathrm{out}} = v_\parallel\big|_{\mathrm{in}} = 0$, the force becomes

$$\boldsymbol{F} = \oint_\Gamma \frac{1}{2}\rho_0 v_\perp^2 \boldsymbol{n} \mathrm{d}\Gamma\,. \tag{S46.}$$

These formulations lead to same expressions presented earlier. It should be noted, however, that for penetrable scatterers, the integral expression above cannot be directly applied, since the relationship between the internal and external acoustic fields is generally not known as a priori. Specifically, the tangential component of the internal velocity. This challenge has been addressed in several classical studies. For instance, under a long-wavelength assumption, the internal field can be approximated using single-particle scattering solutions, allowing the internal contributions to be incorporated into the external field via a rescaling coefficient[10–12].

## S5 The effect of dissipation

### 5.1 The effect of dissipation on the S-matrix

Although our optimization does not explicitly consider loss, the S-matrices are experimentally obtained in lossy conditions owing to the inevitable acoustic loss in reality. Here, we present a theoretical account on the effect of loss and show that it does not hinder our scheme. As mentioned in Appendix G, the loss in our acoustic system is safely considered weak. Such weak loss gives rise to a small imaginary part in the frequency[13]. And the S-matrix can be approximated by

$$\boldsymbol{S}\left(\omega_0 + \frac{\mathrm{i}\delta}{2}\right) \approx \boldsymbol{S}(\omega_0)\left[\boldsymbol{I} - \frac{\delta}{2}\boldsymbol{Q}_\omega(\omega_0)\right], \tag{S47.}$$

where the loss coefficient in our experiment is estimated as $\delta = 0.0032\omega_0 \ll \omega_0$. The operator $\boldsymbol{Q}_\omega$ is the Wigner-Smith time-delay operator.

#### (1) Loss rescales singular values, not phases.

It can be shown that loss primarily affects the amplitudes of the S-matrix while leaving its phases nearly unchanged. Since $\boldsymbol{Q}_\omega$ can be diagonalized by the Wigner-Smith (WS) modes[14,15] as

$$\boldsymbol{Q}_\omega(\omega_0) = \mathbf{W}\overline{\boldsymbol{Q}}\mathbf{W}^\dagger\,, \tag{S48.}$$

where $\overline{\boldsymbol{Q}} = \mathrm{diag}(q_1, q_2, \cdots)$ is real and diagonal. In the same WS basis, the symmetric unitary matrix $\mathbf{S}$ admits a Takagi decomposition[16] of the form

$$\mathbf{S}(\omega_0) = \mathbf{W}^* \overline{\mathbf{S}} \mathbf{W}^\dagger \, , \tag{S49.}$$

with $\overline{\mathbf{S}} = \mathrm{diag}(s_1, s_2, \cdots)$ being a complex and diagonal matrix. By appropriately choosing the phases of $\mathbf{W}$, the phases of $\overline{\mathbf{S}}$ can be removed. The normalized basis is chosen as $\mathbf{W}_{\mathrm{norm}} = \mathbf{W}\overline{\mathbf{S}}^{-1/2}$, in which

$$\mathbf{S}(\omega_0) = \mathbf{W}_{\mathrm{norm}}^* \mathbf{W}_{\mathrm{norm}}^\dagger, \tag{S50.}$$

$$\boldsymbol{Q}_\omega(\omega_0) = \mathbf{W}_{\mathrm{norm}} \overline{\boldsymbol{Q}} \mathbf{W}_{\mathrm{norm}}^\dagger. \tag{S51.}$$

In this basis, the Takagi decomposition of $\mathbf{S}(\omega_0)$ becomes identical to the singular-value decomposition. The shifted S-matrix then becomes

$$\boldsymbol{S}\left(\omega_0 + \frac{\mathrm{i}\delta}{2}\right) \approx \mathbf{W}_{\mathrm{norm}}^* \left(\boldsymbol{I} - \frac{\delta}{2} \overline{\boldsymbol{Q}}\right) \mathbf{W}_{\mathrm{norm}}^\dagger. \tag{S52.}$$

which is also identical to a singular-value decomposition. Compared with the original S-matrix (Eq. (S50)), each singular value is simply scaled by a real factor $s_\mu' = \left(1 - \frac{\delta}{2} q_\mu\right)$, while the singular vectors remain essentially unchanged. Consequently, once the S-matrix is measured, the effect of loss can be removed by replacing its diagonal singular-value matrix with the identity while retaining the left and right singular vectors, thereby recovering the lossless S-matrix.

Moreover, since $\tau = |\mathrm{Tr}\, \boldsymbol{Q}_\omega| = |\mathrm{Tr}\, \overline{\boldsymbol{Q}}| = \left|\sum_\mu q_\mu\right|$ is proportional to the average WS time delay. Configurations with shorter time delay are therefore preferable for mitigating dissipation. Based on these considerations, we select configurations for experimental measurements according to the following criteria. Let $\langle\tau\rangle$ denote the average over 40 realizations, we require $\tau/\langle\tau\rangle < 0.9$. The configurations shown in figure 2 (B1, C1, D1) in the main text are those that exhibit the strongest robustness to loss among the 40 optimization realizations.

**(2) Fidelity analysis.**

The above conclusion can be further quantified by a similarity measure. We use the normalized Frobenius inner product

$$\mathcal{F}(\boldsymbol{A}, \boldsymbol{B}) = \frac{\mathrm{tr}\left(\boldsymbol{A}^\dagger \boldsymbol{B}\right)}{2N}, \tag{S53}$$

where $N$ is the number of channels. $|\mathcal{F}| \in [0,1]$ measures amplitude correlation, $\arg \mathcal{F}$ captures any global phase difference, and two matrices differing only by a global phase are naturally identified. For $\boldsymbol{A} = \boldsymbol{S}(\omega_0 + \mathrm{i}\delta/2)$ and $\boldsymbol{B} = \boldsymbol{S}(\omega_0)$

$$\mathcal{F}\left[\boldsymbol{S}\left(\omega_0 + \frac{\mathrm{i}\delta}{2}\right), \boldsymbol{S}(\omega_0)\right] \approx \frac{\operatorname{tr}\left(\boldsymbol{I} - \frac{\delta}{2}\boldsymbol{Q}_\omega\right)}{2N} + O(\delta^2) = 1 - \frac{\delta}{4N}\operatorname{tr}\boldsymbol{Q}_\omega + O(\delta^2), \tag{S54}$$

where we used $\operatorname{tr}\boldsymbol{I} = 2N$, and the hermiticity of $\boldsymbol{Q}_\omega$. The deviation is real and linear in $\delta$, confirming that loss introduces no global phase shift. Once the singular values are normalized to remove the amplitude scaling, the similarity measure becomes

$$\tilde{\mathcal{F}}\left[\boldsymbol{S}\left(\omega_0 + \frac{\mathrm{i}\delta}{2}\right), \boldsymbol{S}(\omega_0)\right] \approx \frac{\operatorname{tr}\left(\boldsymbol{I} - \frac{\delta}{2}\boldsymbol{Q}_\omega\right)}{\sqrt{\operatorname{tr}\boldsymbol{I}}\sqrt{\operatorname{tr}\left(\boldsymbol{I} - \delta\boldsymbol{Q}_\omega + \frac{\delta^2}{4}\boldsymbol{Q}_\omega^\dagger\boldsymbol{Q}_\omega\right)}} \approx 1 + O(\delta^2). \tag{S55}$$

We remark that the first-order correction vanishes and the residual deviation is quadratic in $\delta$. It means that the fidelity is insensitive to small $\delta$

We have further performed a parameter sweep of $\delta$ in COMSOL for the generator $\sigma_1$ at 3200 Hz. Figure S5(A) shows the unnormalized inner product $\mathcal{F}$: it decreases linearly with $\delta$, in agreement with Eq. (S54). Figure S5(B) shows the normalized fidelity $\tilde{\mathcal{F}}$ with singular-value-normalized S-matrices,where a quadratic dependence on $\delta$ is clearly seen. The loss value marked by red and blue circles in Figure S5 are obtained by fitting experimental data with COMSOL simulations.

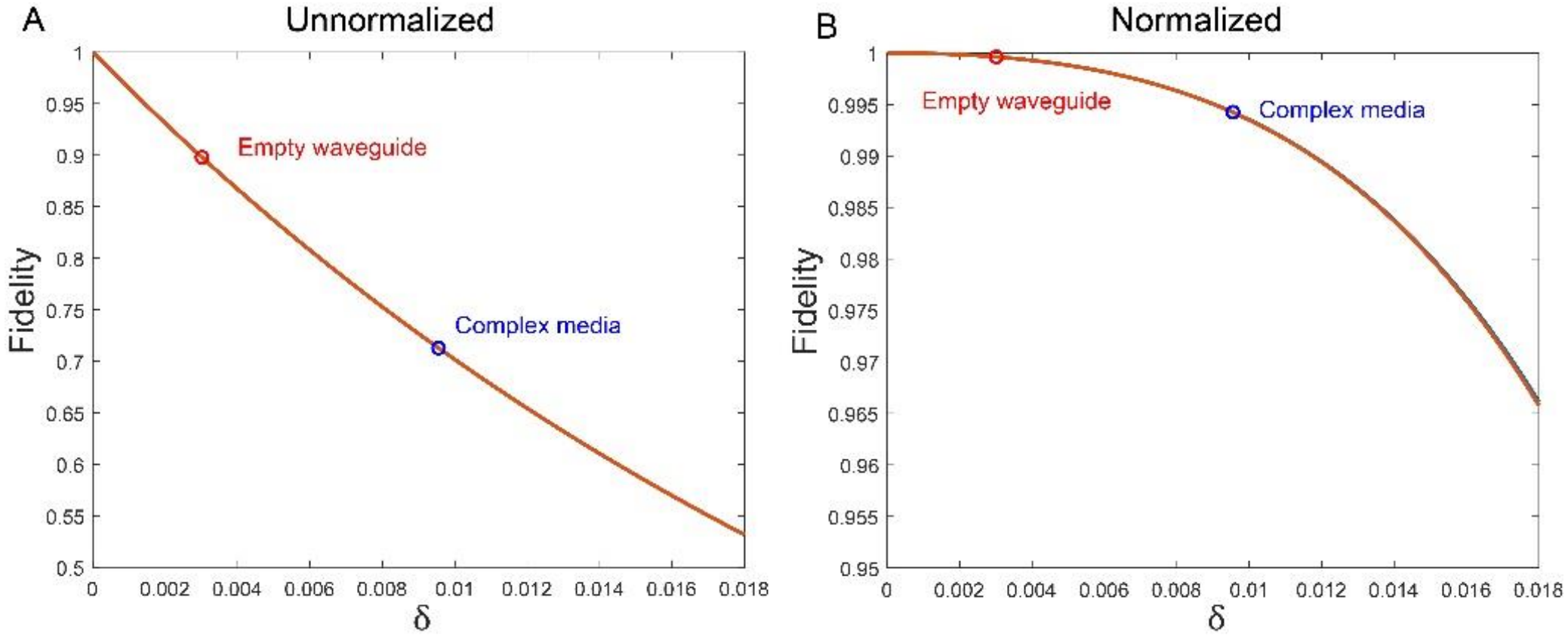

**Figure S5**. The effect of loss on the S-matrix for the generator $\sigma_1$ at 3200 Hz. (A) The unnormalized inner product $\mathcal{F}$ versus linear decay $\delta$. (B) The normalized fidelity $\tilde{\mathcal{F}}$ after singular-value normalization shows a quadratic relation with respect to $\delta$.

### 5.2 Experimental characterization of intrinsic loss

Figure S6 shows the measured S-matrix of a 1-m-long air-filled waveguide at 3200 Hz. In the ideal lossless case, the reflection matrix is zero and the transmission matrix is diagonal. The measured phases agree well with theoretical prediction (Table S1), and the amplitudes reduces to approximately 0.9 of the ideal value.

Figure S7 compares the unnormalized experimental transmission matrices of the three optimized generators with their theoretical counterparts. Despite the presence of loss, the phase structure is excellently preserved across all three configurations. The dominant mode in each channel accounts for up to ~90% of the relative amplitude. The energy transmission is approximately 74%; after accounting for the 90% baseline of the empty waveguide, the net transmission attributable to the scattering medium is ~80%. The slightly increased loss relative to the empty waveguide is expected, as the scatterers introduce additional surface area and longer propagation paths.

In summary, under the weak-loss conditions of our experiment, loss rescales the singular values of transmission coefficients but does not affect the phases. Singular-value normalization therefore recovers the ideal lossless S-matrix with high accuracy.

**Table S1**: The measured vs. theoretical phases of the T-matrix diagonals of a 1m air-filled waveguide at 3200 Hz.

| / | Arg $t_{11}$ | Arg $t_{22}$ | Arg $t_{33}$ | Arg $t_{44}$ |
|---|---|---|---|---|
| Exp.(unnormalized) | $-100.6°$ | $22.90°$ | $67.4°$ | $-157.0°$ |
| Lossless air | $-99.1°$ | $24.4°$ | $67.6°$ | $-164.6°$ |

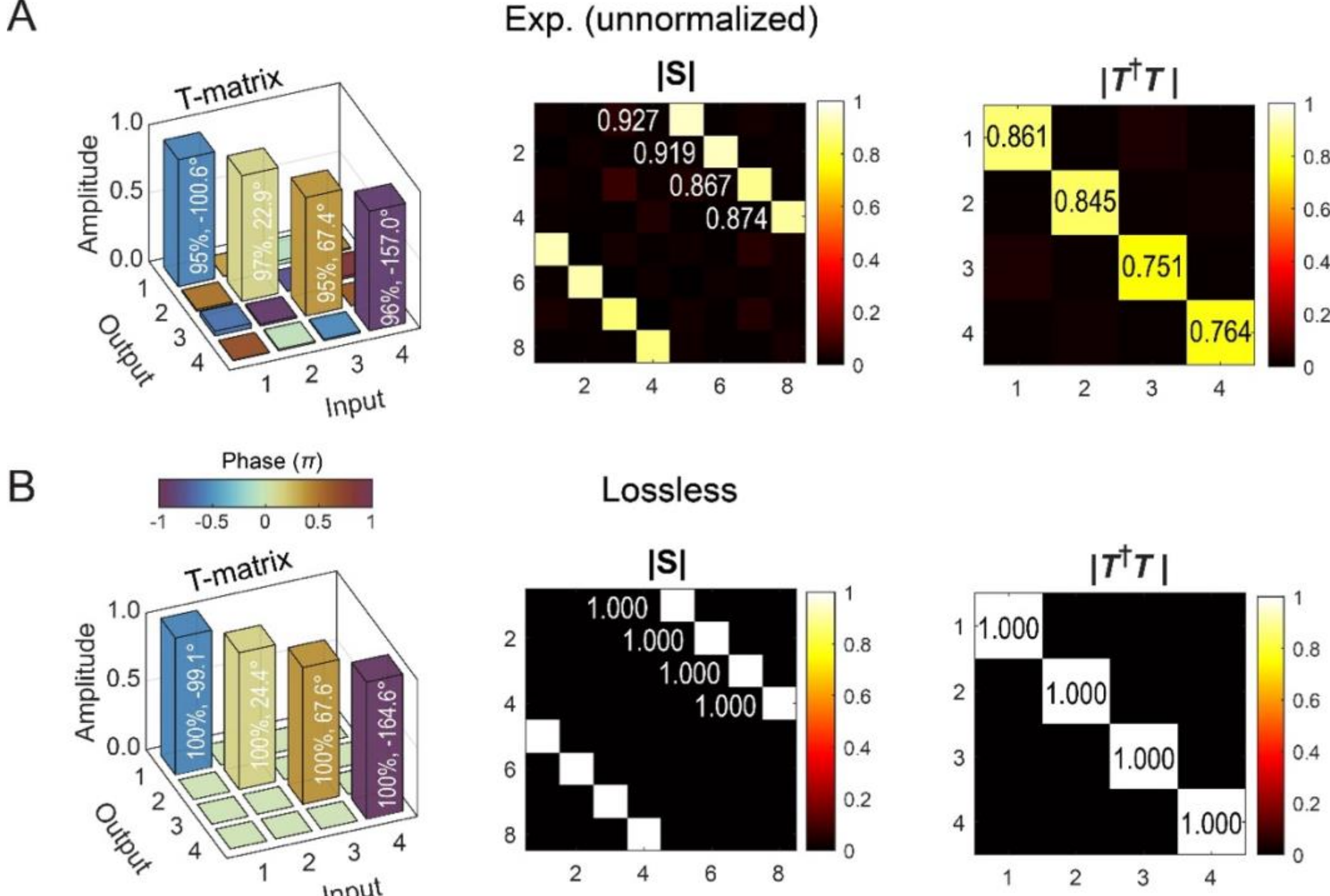

**Figure S6**. The S-matrices of an air-filled waveguide, one meter in length, at 3200 Hz. (A) The experimentally measured, unnormalized results. The left column: the transmission matrix, where the bar heights represent the amplitudes and the colormap represents the phase. The transmission amplitudes and phases of the diagonals are marked. The middle column: amplitude of the S-matrix. The right column: amplitude of the energy transmission matrix, with annotated numbers indicating the element amplitudes. (B) The lossless results computed by theory.

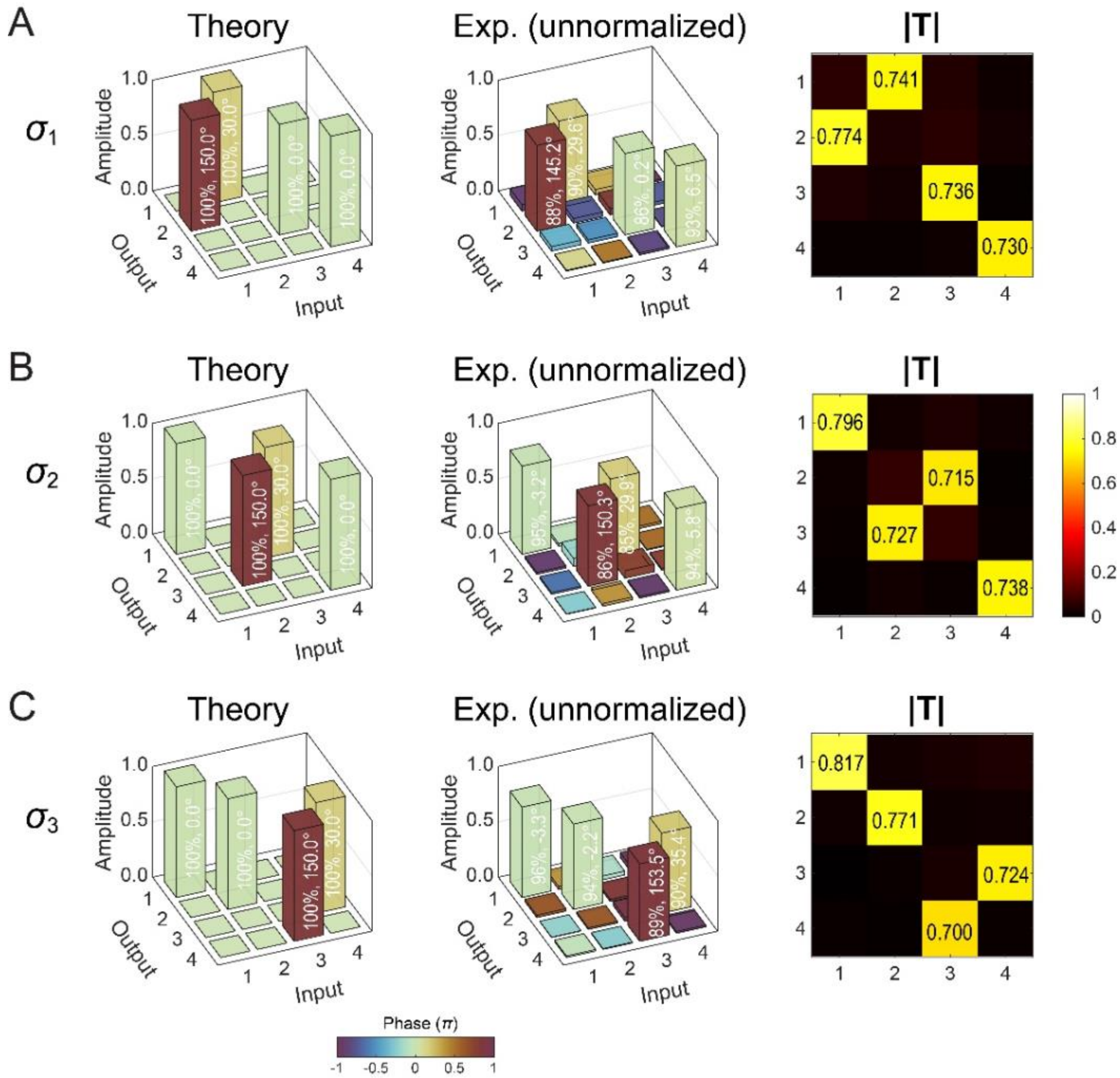

**Figure S7**. The unnormalized and theoretical T-matrices for the three generators at 3200 Hz (A) Generator $\sigma_1$: theoretical (left), experimental (middle), and amplitude of the experimental T-matrix (right). (B, C) Generators $\sigma_2$ and $\sigma_3$, respectively.

### 5.3 The acoustic Wigner-Smith time-delay operator

We now derive the expression of the acoustic Wigner-Smith time-delay operator (WSTO). The approach follows closely the derivation of WSTO for electromagnetic waves[14,17]. To maintain consistency with the form of the GWSO, we define

$$\boldsymbol{Q}_\omega = -\mathrm{i}\boldsymbol{S}^{-1}(\omega)\frac{\partial \boldsymbol{S}(\omega)}{\partial \omega} = -\mathrm{i}\boldsymbol{S}^{\dagger}\frac{\partial \boldsymbol{S}(\omega)}{\partial \omega}. \tag{S56}$$

Time dependence is taken as $e^{\mathrm{i}\omega t}$. The frequency-domain linear acoustic equations are

$$-\nabla P = \mathrm{i}\omega\rho\boldsymbol{v}, \tag{S57}$$

$$-\nabla \cdot \boldsymbol{v} = \mathrm{i}\omega\kappa^{-1}P. \tag{S58}$$

All material parameters and $\omega$ are assumed real unless stated otherwise. By analogy with the electromagnetic wave equation[14,17], we obtain

$$\boldsymbol{Q}_\omega = \frac{1}{2\omega}\left[\mathrm{i}\left(\boldsymbol{D}\boldsymbol{S} - \boldsymbol{S}^\dagger\boldsymbol{D}\right) - \boldsymbol{E}\right]. \tag{S59}$$

Herein, the first term involves the S-matrix and the dispersion matrix $\boldsymbol{D}$, which is diagonal with entries $D_{mm} = \frac{(k_m^y)^2}{(k_m^x)^2}$. The second term $\boldsymbol{E}$ is a volume integral involving the acoustic energy density in the scattering region $\Omega$, with entries $E_{nm} = \int_\Omega (\nabla P_n^* \cdot \nabla P_m + k_0^2 P_n^* P_m)\mathrm{d}\Omega$. Thus, the time delay is

$$\tau = |\mathrm{tr}\,\boldsymbol{Q}_\omega| = \frac{1}{2\omega}\sum_{n=1}^{N}\int_\Omega (|\nabla P_n|^2 + k_0^2|P_n|^2)\mathrm{d}\Omega\,. \tag{S60}$$

$N$ is the total number of channels.

**S6 Bandwidth analysis**

As shown in Appendix H of the main text, the fidelity decays quadratically with small frequency detuning $\Delta\omega$. To validate this, we simulated the disordered medium corresponding for generator $\sigma_1$. Sweeping the excitation from 3000 to 3400 Hz and comparing each S-matrix against the one at 3200 Hz, we obtain the results shown in Figure S8. The fidelity $|\mathcal{F}|$ stays above 0.90 over 3119-3279 Hz (160 Hz bandwidth) and above 0.95 over 3139-3257 Hz (118 Hz bandwidth). The four frequencies marked in Figure S8(A), i.e., 3200, 3170, 3139, 3119 Hz, correspond to 100%, 99%, 95% and 90% fidelity respectively; their field patterns are shown in Figure S8(B-E). These results confirm that the quadratic dependence on frequency shift ensures the single-frequency optimized solution remains accurate over a practically useful bandwidth.

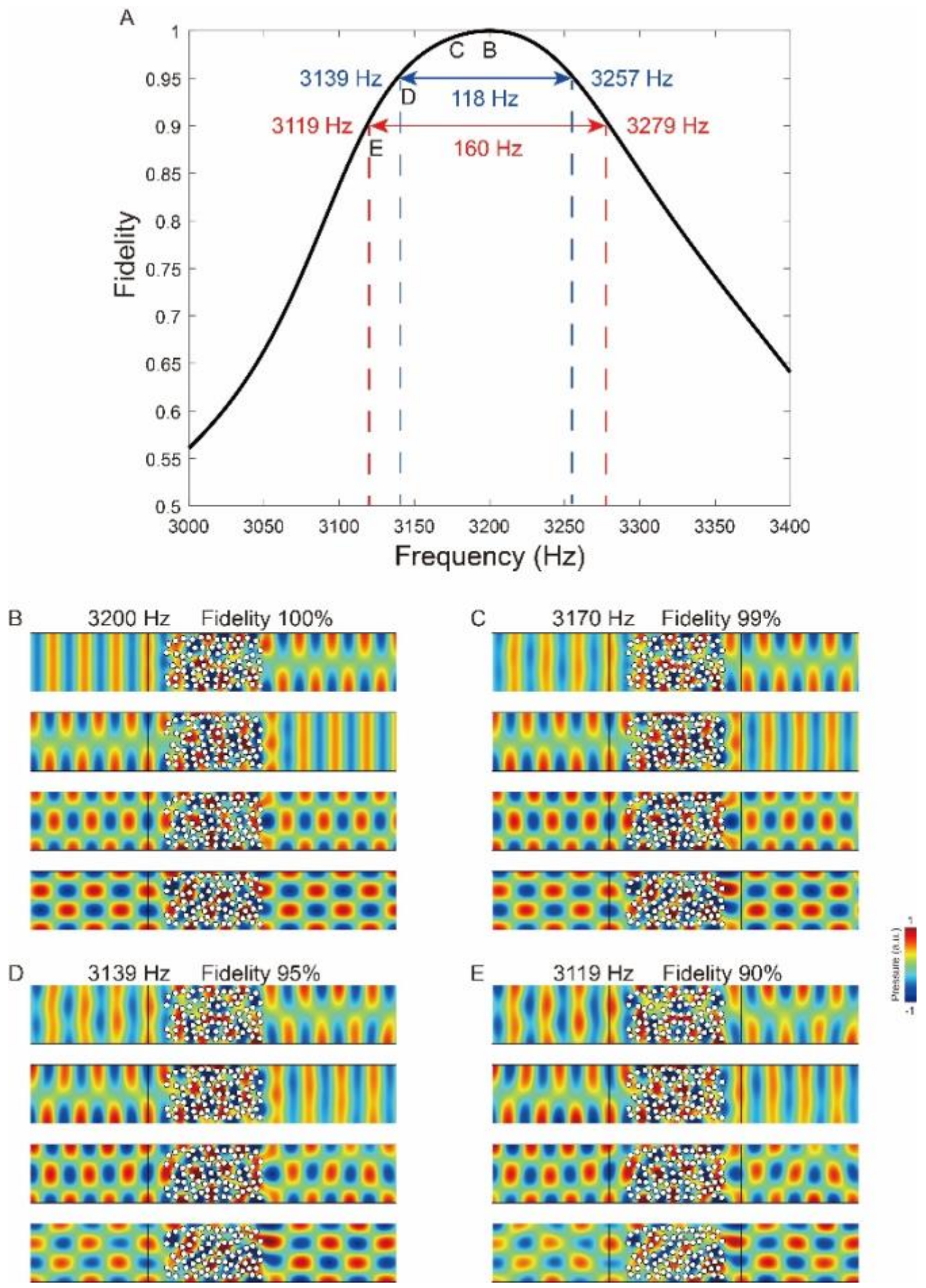

**Figure S8.** Frequency bandwidth for generator $\sigma_1$. (A) Fidelity versus frequency, reference at 3200 Hz. (B-E) Braiding field patterns at 3200 Hz, 3170 Hz, 3139 Hz, and 3119 Hz.

### S7 Errors in scatterer positions

As shown in Appendix H, small errors in scatterer positions lead to a quadratic fidelity decay, implying inherent robustness against small positioning errors. To assess this in practice, we consider two tolerance-related contributions: (1) manual placement using a printed A4 paper template (line width $\sim$0.25 mm, typical placement error <0.5 mm); (2) fabrication errors of the cylinders produced by CNC machining (easily achieving $\pm$0.01$\sim$0.1 mm). Therefore, the major source of errors is induced by the placement of the cylinders by hand. The machining fabrication is negligible.

We generated 1000 random realizations with displacement magnitudes uniformly distributed in the range of 0$\sim$0.5 mm (average ~0.25 mm, i.e., ~5 % of the cylinder radius), and directions uniformly distributed in $[0,2\pi)$. The results are shown in Figure S9. The fidelity values exceed 0.98 (values above 0.95 are typically regarded as excellent); the red line marks the mean fidelity of 99.29 %, and the corresponding field pattern confirms that even manual placement yields a high-quality result.

We remark that the good agreement with the experimental results and theoretical predictions also confirms the tolerance of errors in scatterer positions: the cylinders were placed by hand with simple visual guidance.

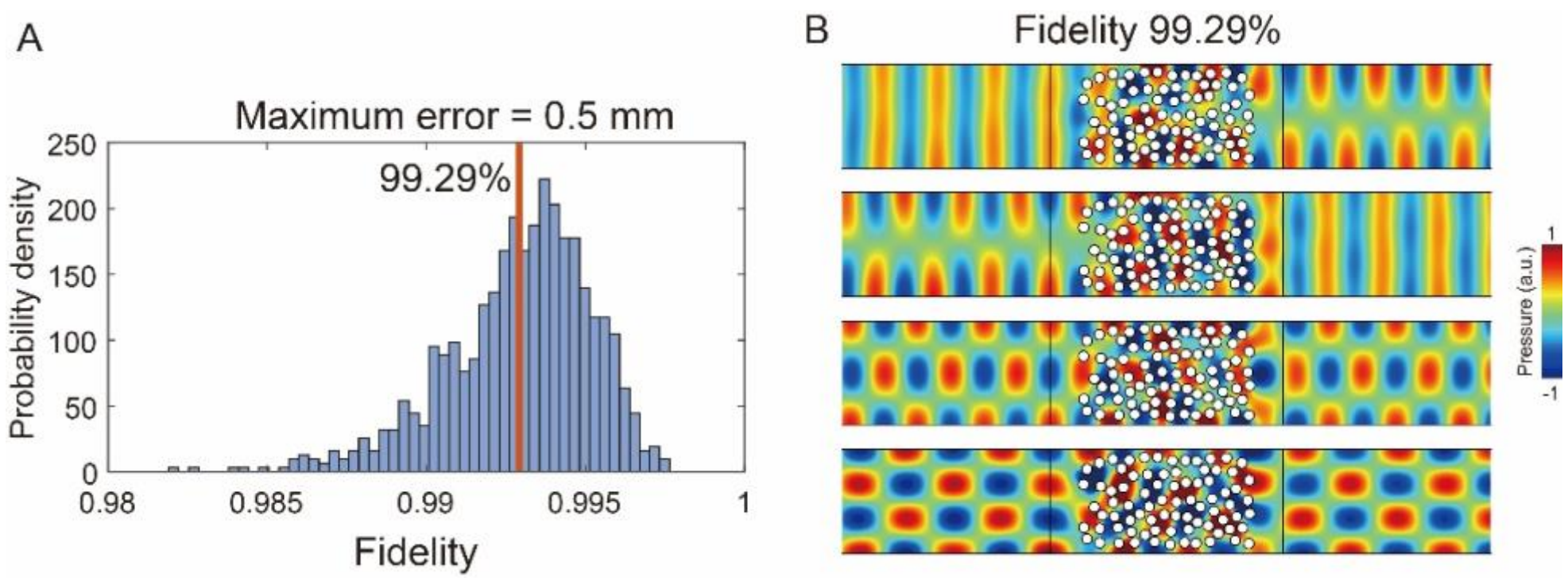

**Figure S9.** Positioning errors analysis for the generator $\sigma_1$. (A) The fidelity histogram for 1000 realizations with a maximum error of 0.5 mm. The mean fidelity is 99.29%. (B) The corresponding field patterns.